\documentclass[10pt,twocolumn,letterpaper]{article}
\usepackage[accsupp]{axessibility}
\usepackage{cvpr}              

\usepackage{makecell}
\usepackage{tikz}
\usepackage{booktabs}          

\usepackage{keyval}
\definecolor{cvprblue}{rgb}{0.21,0.49,0.74}
\usepackage[pagebackref,breaklinks,colorlinks,allcolors=cvprblue]{hyperref}
\usepackage{wrapfig}
\usepackage{algorithm,algorithmic}

\def\confName{CVPR}
\def\confYear{2026}

\title{SmokeSVD: Smoke Reconstruction from A Single View via Progressive Novel View Synthesis and Refinement with Diffusion Models}

\author{
	\begin{tabular}[t]{cccc}
		Chen Li$^{1\dag}$ & Shanshan Dong$^{2\dag}$ & Sheng Qiu$^{2*}$ & Jianmin Han$^{2}$ \\[4pt]
		Yibo Zhao$^{1}$ & Zan Gao$^{1}$ & Taku Komura$^{3}$ & Kemeng Huang$^{3}$
	\end{tabular}
	\\
	{
		$^1$ Tianjin University of Technology \quad
		$^2$ Zhejiang Normal University \quad
		$^3$ University of Hong Kong
	}
}

\begin{document}

\twocolumn[{%
\renewcommand\twocolumn[1][]{#1}%
\maketitle
\centering
\includegraphics[width=1.0\linewidth]{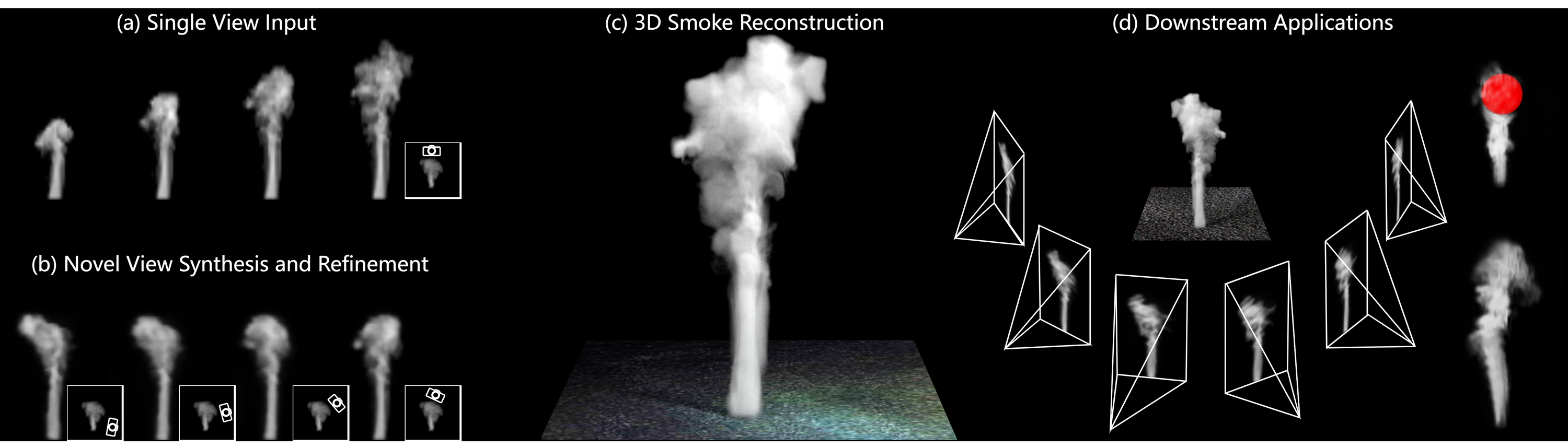}
\captionof{figure}{By leveraging physics-aware diffusion and refinement modules, our method progressively performs novel view synthesis (b) and 3D reconstruction (c) from a single-view input (a). When applied to downstream applications (d), our approach enables flexible novel view generation, re-simulation, and artist-driven control.}
\label{fig:teaser}
}]
\renewcommand{\thefootnote}{\fnsymbol{footnote}}
\footnotetext[1]{Corresponding Author. 	$^\dag$Equal contributions. }
\renewcommand{\thefootnote}{\arabic{footnote}}

\begin{abstract}
Reconstructing dynamic fluids from sparse views is a long-standing and challenging problem, due to the severe lack of 3D information from insufficient view coverage. While several pioneering approaches have attempted to address this issue using differentiable rendering or novel view synthesis, they are often limited by time-consuming optimization under ill-posed conditions. We propose SmokeSVD, an efficient and effective framework to progressively reconstruct dynamic smoke from a single video by integrating the generative capabilities of diffusion models with physically guided consistency optimization. Specifically, we first propose a physically guided side-view synthesizer based on diffusion models, which explicitly incorporates velocity field constraints to generate spatio-temporally consistent side-view images frame by frame, significantly alleviating the ill-posedness of single-view reconstruction. Subsequently, we iteratively refine novel-view images and reconstruct 3D density fields through a progressive multi-stage process that renders and enhances images from increasing viewing angles, generating high-quality multi-view sequences. Finally, we estimate fine-grained density and velocity fields via differentiable advection by leveraging the Navier-Stokes equations. Our approach supports re-simulation and downstream applications while achieving superior reconstruction quality and computational efficiency compared to state-of-the-art methods.
\end{abstract}

\section{Introduction}
\label{sec:introduction}
Smoke reconstruction and motion estimation from RGB videos has always been an important issue in a wide range of fields, including computer graphics and vision~\cite{wang2024physics}, atmospheric physics~\cite{carrico2010water}, optics~\cite{han2025three}, medicine~\cite{chen2019smokegcn}.
Despite the rapid development of dynamic radiance fields, it is cumbersome and sometimes impractical for non-specialist to capture multi-view images of smoke phenomena in non-laboratory environments, impeding the widespread applications of relevant techniques, therefore, efficiently reconstructing and understanding smoke phenomena from highly sparse captured images~\cite{liu2011modeling} is of great value. 

Existing solutions~\cite{gregson2014capture, okabe2015fluid, eckert2018coupled, zang2020tomofluid} for sparse-view fluid capture integrate physically-based and geometric priors but are time-consuming. For single-view reconstruction, \citet{franz2021global} introduced physical priors via differentiable rendering, but remains computationally expensive. Recent works~\cite{gao2025fluidnexus, chen2024cascade} employ diffusion models to generate novel view videos, alleviating the ill-posed problem. However, combining multi-view diffusion models with sparse-view reconstruction faces two challenges: (1) limited multi-view consistency, where diffusion models produce low-quality inconsistent images~\cite{chen2024cascade, zou2024triplane}, and (2) insufficient incorporation of physical priors to guide generative models for complex smoke dynamics and external inflows.

In this paper, we propose SmokeSVD for efficient high-quality smoke reconstruction from single-view video.
Inspired by recent 3D generation work~\cite{zou2024triplane}, we first synthesize side-view sequences from front-view input using diffusion models guided by spatial and temporal priors. 
We then progressively generate novel views from near to far. Each iteration reconstructs a coarse 3D density field, then refines novel views using differentiable rendering and UNet3+~\cite{huang2020unet} for visual fidelity and temporal coherence. Finally, we reconstruct fine-grained density and velocity fields, and infer inflow states to support 
downstream applications.

Unlike recent sparse-view methods~\cite{gao2025fluidnexus} that first generate multi-view images then reconstruct 3D, leading to shape-appearance ambiguity from insufficient consistency, we advocate a multi-stage strategy cyclically utilizing 2D diffusion synthesis, spatio-temporal refinement, and coarse/fine-grained 3D reconstruction. This exploits both high-quality 2D diffusion outputs and 3D volumetric consistency. Our progressive generation is guided by multi-view consistent optimization for temporally coherent sequences with minimal computation. Thus, SmokeSVD outperforms state-of-the-art in both quality and efficiency.

Our contributions are summarized as follows:
\begin{itemize}
	\item We propose a novel and efficient smoke reconstruction framework from a single view by incorporating multi-stage 2D novel view synthesizer/refinement and coarse/fine-grained 3D reconstruction. The proposed framework allows us to rapidly infer velocity field and dynamic inflow states, supporting re-simulation of the input phenomena, or generation of new visual effects.

	\item We propose a method to synthesize a visually plausible side view image sequences based on front view sequences using the diffusion model. To guarantee reasonable smoke motion, we incorporate 3D predicted density and velocity fields as physical guidance into the denoising process for enhancing temporal consistency and producing physically-plausible smoke motion.  
	
	\item We present a novel view refinement approach to progressively produce high-quality and consistent multi-view image sequences by injecting multiple view information and coarse 3D density field. Compared to direct multi-view diffusion models, our refinement approach achieves a better balance between computational efficiency and reconstruction robustness.
\end{itemize}

\section{Related Work}
\label{sec:related_work}

\paragraph{Fluid Simulation and Reconstruction.}
Physically-based fluid simulation has a long history in computer graphics~\cite{stam1993turbulent, zhou2024eulerian, liu2024dual, tu2024unified, wang2024eulerian,zhang2014hybrid}. Please refer to~\cite{wang2024physics} for a comprehensive survey. 
As the inverse problem, fluid reconstruction are challenging~\cite{xie2024dynamic, xie2024fluid}. Conventional methods rely on specialized devices (e.g., Schlieren photography~\cite{atcheson2008time}, structured light~\cite{gu2012compressive}, light field probes~\cite{ji2013reconstructing}), or passive techniques~\cite{xiong2017rainbow, schneiders2016dense}. 
\citet{gregson2014capture} coupled fluid simulation into flow tracking to reconstruct temporally coherent velocity fields. Similarly, \citet{eckert2018coupled, eckert2019scalarflow} adopted specific simulator components to infer unknown physical quantities. 

Recently, neural rendering has gained attention in fluid reconstruction~\cite{qiu2024neusmoke}. PINF~\cite{chu2022physics} introduces a hybrid representation for dynamic fluid scenes with static obstacles. HyFluid~\cite{yu2023inferring} advocates hybrid neural fields to jointly infer density and velocity from multi-view videos. PICT~\cite{wang2024physics} proposes a neural characteristic trajectory field with spatial-temporal NeRF. However, neural rendering faces challenges capturing high-frequency information from sparse views, often producing over-smooth results.

For single-view reconstruction, GlobTrans~\cite{franz2021global} employs strict differentiable physical priors. \citet{franz2023learning} applied central constraints with differentiable rendering to ensure smoke appearance in novel views. FluidNexus~\cite{gao2025fluidnexus} reconstructs smoke by synthesizing multi-view videos. However, consistency issues may exist among viewpoints generated by~\cite{liu2023zero}. Our method alleviates the ill-posed problem by generating side-view sequences and uses progressive refinement to ensure multi-view consistency.

\paragraph{Novel View Synthesis with 2D Diffusion Models.}
Since~\cite{ho2020denoising}, diffusion models have been widely applied to multiple domains~\cite{xing2024survey,ho2022video,huang2023make,yu2024texgen,shi2024interactive}. Through implicit representations~\cite{rombach2022high} and sampling techniques~\cite{zhang2023gddim}, diffusion models achieve high quality and speed. Several studies have applied diffusion models to novel view synthesis~\cite{liu2023zero, watson2023novel, tseng2023consistent}. Zero-1-to-3~\cite{liu2023zero} and 3DiM~\cite{watson2023novel} concatenate conditional information as model inputs, while pose-guided diffusion uses cross-attention. However, end-to-end generation may lack consistency across viewpoints. To improve consistency, multiple works~\cite{shi2023zero123plus, shi2023mvdream, weng2023consistent123, yang2024consistnet, kwak2024vivid} have been proposed. Zero123++~\cite{shi2023zero123plus} learns joint distribution by combining multi-view images into one. MVDream~\cite{shi2023mvdream} enhances consistency via 3D self-attention and MLPs for camera information. Consistent123~\cite{weng2023consistent123} introduces cross-view and shared self-attention for structural consistency. ConsistNet~\cite{yang2024consistnet} back-projects features into 3D space using multi-view geometry. ViVid-1-to-3~\cite{kwak2024vivid} reformulates it as video generation, introducing video diffusion priors.
However, existing methods cannot be directly applied to smoke due to its complex physical properties.

\section{Method}
\label{sec:method}

\subsection{Overview}
\label{subsec:framework}

\begin{figure*}[htb]
	\centering
	\includegraphics[width=0.85\linewidth]{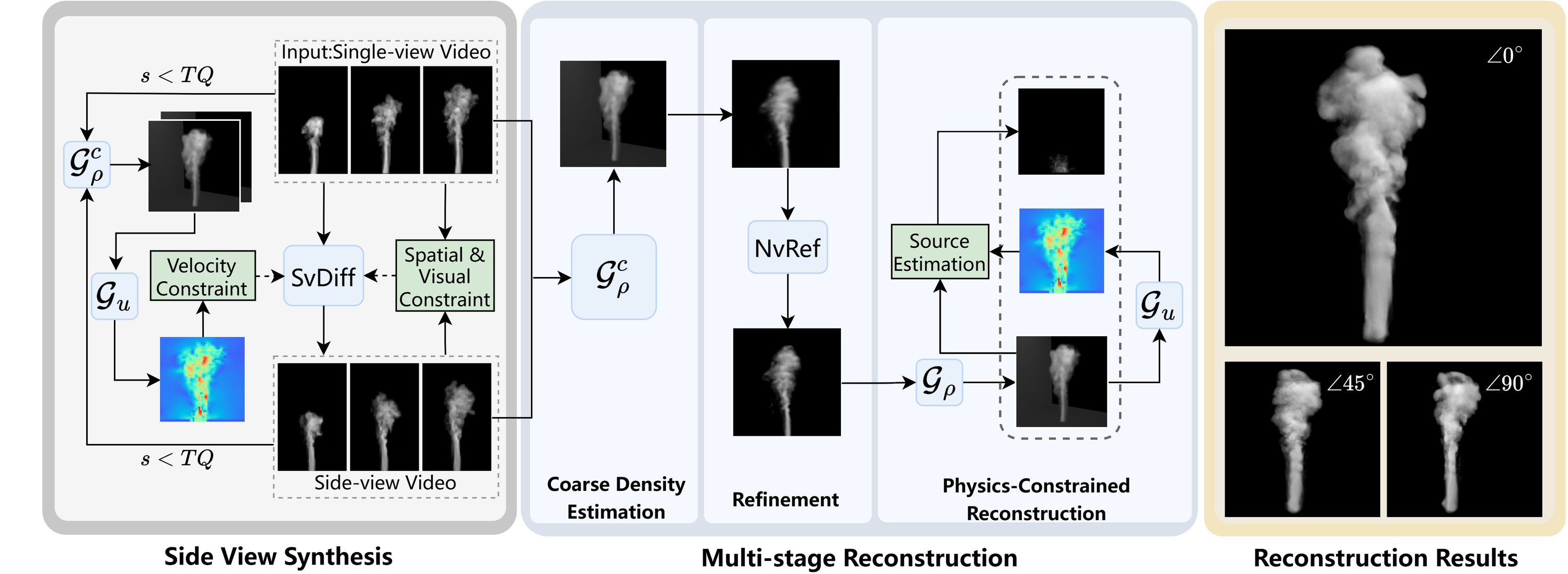}
	\caption{Overview of SmokeSVD. We categorize view angles into three types: input as front view ($\alpha = \angle0^\circ$), synthesized images from $\text{SvDiff}$ as side view ($\alpha = \angle90^\circ$), and all others as novel views. Given front-view input, we first synthesize side-view sequences guided by spatial, visual, and velocity constraints via density and velocity reconstruction. We then iteratively estimate coarse 3D density and refine novel-view sequences, progressively introducing views from near to far. Our pipeline outputs 3D density, velocity fields, and dynamic inflow. Physical priors guide both $\text{SvDiff}$ and $\text{NvRef}$ modules for physically accurate and visually realistic results.}
	\label{fig:overview}
\end{figure*}

Our pipeline is illustrated in Fig.~\ref{fig:overview}. Given a single-view video of $T$ frames, we treat it as the front-view sequence $w^t_{\angle0^\circ}$, where $t$ is the frame number and $\alpha=\angle0^\circ$ denotes the offset angle from front view. 
We propose a side-view synthesizer $\text{SvDiff}$ based on diffusion models to synthesize side-view video $w^t_{p,\angle90^\circ}$ from $w^t_{\angle0^\circ}$ with reasonable spatial distribution, temporal evolution and appearance.
Then, a coarse-grained density generator $\mathcal{G}^c_{\rho}$ generates a rough 3D density field $\rho_{r,c}$ from $w^t_{\angle0^\circ}$ and $w^t_{p,\angle90^\circ}$. 
We progressively rotate the camera along the horizontal plane to render novel view images (e.g., $w^t_{r,\angle45^\circ}$, $w^t_{r,\angle135^\circ}$), and refine them frame by frame with novel view refinement module $\text{NvRef}$. Benefiting from 3D spatial distribution constraint from $\rho_c$ and temporal-spatial correlation from UNet3+, $\text{NvRef}$ produces multi-view consistent images. 
With multiple views, we employ a fine-grained density generator $\mathcal{G}^f_{\rho}$ to reconstruct high-quality density field $\rho_{r,f}$, jointly estimating velocity fields $\mathbf{u}$ and inflow states $\rho_{in}$ via differentiable advection operator $\mathcal{A}$, ensuring reconstruction satisfies long-term physical constraints.
Finally, we can re-simulate the input smoke and support downstream applications, e.g., novel view synthesis, artist control.

\subsection{Physically-Aware Side-View Synthesizer}
\label{subsec:synthesizer}

While substantial progress has been made in generalizable novel-view synthesis, most approaches lack effective physically-aware priors for complex volumetric phenomena. Smoke poses unique challenges due to its semi-transparent appearance and complex dynamics. First, ensuring spatiotemporal consistency across synthetic sequences is difficult, as current methods often produce visual artifacts including temporal flickering and motion incoherence. Second, maintaining cross-view consistency between input frontal and generated side views requires sophisticated modeling of shared volumetric properties, as both views represent different projections of the same 3D volume with consistent spatial distributions and appearance. 

We incorporate physical and visual priors into our side-view synthesizer $\text{SvDiff}$ to address these challenges. $\text{SvDiff}$ extends image generation diffusion models~\cite{ho2020denoising} to handle smoke sequences frame-by-frame for temporal coherence. Inspired by classifier-free guidance~\cite{ho2021classifier}, we use side-view images of two previous frames $w^{t-1}_{\angle90^\circ}, w^{t-2}_{\angle90^\circ}$ and current front-view image $w^t_{\angle0^\circ}$ as condition to train $\text{SvDiff}$:
\begin{equation}
	c^t=w^t_{\angle0^\circ} \oplus w^{t-1}_{\angle90^\circ} \oplus w^{t-2}_{\angle90^\circ},
	\label{eq:p_old}
\end{equation}
where $\oplus$ denotes concatenation. For initial frames ($t<2$), we train another synthesizer with condition $c^0=w^0_{\angle0^\circ} \oplus w^1_{\angle0^\circ}$. 
$\text{SvDiff}$ is trained by minimizing: 
\begin{equation}
	\mathcal{L}_{noise} = 
	\|\epsilon-\epsilon_\theta (w^{t}_{\angle90^\circ},c^t,s) \|^2.
	\label{eq:ourLoss}
\end{equation}
During training, $\text{SvDiff}$ synthesizes side-view image $w^{t}_{\angle90^\circ}$ from ground truth side-view images $w^{t-1}_{\angle90^\circ}, w^{t-2}_{\angle90^\circ}$. However, during inference, $\text{SvDiff}$ uses previously synthesized frames as input conditions, which progressively accumulates errors over time.

\begin{figure*}[ht]
	\centering
	\includegraphics[width=0.8\linewidth]{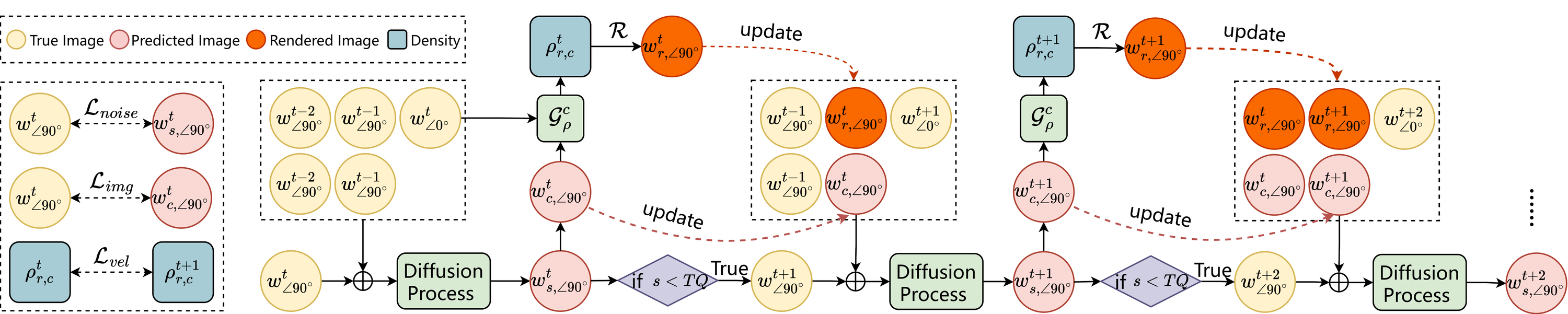}
	\caption{Frame-by-frame training of the side-view synthesizer via feature fusion of adjacent frames. In the forward diffusion process, a clean image $w_{c,\angle 90^\circ}$ is estimated from the noisy image $w_{s,\angle t,90^\circ}$ , and this estimated clean image serves as one of the conditional images for the next forward diffusion process. The figure demonstrates the forward diffusion training process for three consecutive frames.}
	\label{fig:dm}
\end{figure*}

To reduce accumulated error and ensure long-term stability, we propose a multi-frame training scheme enabling $\text{SvDiff}$ to learn from both historically generated and rendered images of reconstructed density fields, as shown in Fig.~\ref{fig:dm}. We re-formulate Eq.~\ref{eq:p_old} as 
$c^t=w^t_{\angle0^\circ} \oplus  w^{t-1}_{c,\angle90^\circ }\oplus  w^{t-1}_{r,\angle90^\circ} \oplus  w^{t-2}_{c,\angle90^\circ} \oplus  w^{t-1}_{r,\angle90^\circ}$, where $ w_c$ is the synthesized side-view image from $\text{SvDiff}$. Since diffusion training predicts noise from the forward process, in multi-frame training we estimate generated images from noise. Based on Eq~\ref{eq:ourLoss}, the estimated clean image is:
\begin{equation}
	w_{\angle 90^\circ} \approx w_{c,\angle 90^\circ}=\frac{w_{s,\angle 90^\circ}-\sqrt{1-\Bar{\alpha}_s}\epsilon_\theta}{\sqrt{\Bar{\alpha}_s}},
	\label{eq:x_0}
\end{equation}
where $w_{s,\angle \alpha}$ denotes a noisy image at diffusion step $s$ and viewpoint $\alpha$. When $s$ is not labeled, it defaults to zero, indicating a clean image.

Unlike traditional diffusion models performing one forward process per batch, our multi-frame training performs multiple forward processes per batch. In each forward process, $\text{SvDiff}$ estimates a clean image from the noisy image and uses it as condition for the next forward process. Through multiple forward diffusions, $\text{SvDiff}$ learns from historically generated information, improving long-term stability. 

To incorporate physical and visual priors and guide $\text{SvDiff}$ toward physically faithful results, we introduce a guidance module imposing targeted constraints on denoising.
We set a threshold $TQ$ to determine when the guidance is applied: if $s \geq TQ$, the noise level is too high to extract meaningful physical information between consecutive frames, so the guidance is disabled; otherwise, the guidance module is activated and incorporated into the training objective. Specifically, the guidance consists of three loss terms: visual, velocity and spatial constraints, that collectively steer the model toward more accurate and realistic generation.

\textit{Visual Constraint.} We use $L_2$ loss to measure difference between predicted clean image $\hat{x}_0^i$ and ground truth $x_0^i$, where $i$ denotes the multi-frame training iteration index. This loss $\mathcal{L}_{img} = \|x_0^i - \hat{x}_0^i\|^2$ penalizes pixel-wise discrepancies, ensuring high fidelity.

\textit{Velocity Constraint.} 
To further ensure physically plausible smoke dynamics over time, we introduce velocity constraints between consecutive frames, penalizing both the divergence and abrupt changes in the velocity fields. To infer the 3D velocity field from 2D images, we first use a density generator $\mathcal{G}\rho$ (see Sec.~\ref{subsec:refinement}) to reconstruct a coarse-grained 3D density field $\rho^i_{r,c}$ from the input front-view image and the predicted clean side-view image, defined as $\rho^i_{r,c} = \mathcal{G}_\rho(w^{i+t}_{\angle 0^\circ}, w^{i+t}_{c,\angle 90^\circ})$. Based on these reconstructed density fields from consecutive frames, we then employ a velocity generator $\mathcal{G}_u$ (see Sec. C.5 in supplementary) to estimate the velocity field as $\mathbf{u}^{i-1} = \mathcal{G}_u(\rho^{i-1}, \rho^i_{r,c})$. The velocity constraint consists of two terms:
\begin{equation}
    \label{eq:L_vel}
    \mathcal{L}_{vel} = \|\nabla \cdot \mathbf{u}^{i-1}\|^2 + \|\nabla \mathbf{u}^{i-1}\|^2,
\end{equation}
where the first term enforces incompressibility and the second promotes smoothness, preventing temporal artifacts.

\textit{Spatial Constraint.} 
To ensure that the generated side-view image $w_{c,\angle90^\circ}$ is consistent with the input image $w_{\angle0^\circ}$ in spatial distribution, we design a spatial distribution constraint based on the estimated clean image. The purpose of this loss term is to make \text{SvDiff} more attentive to the spatial distribution differences between $ {w}_{c,\angle90^\circ}$ and $w_{\angle0^\circ}$, thereby guiding \text{SvDiff} to generate features that are closer to ground truth:
\begin{equation}
	\mathcal{L}_{sp} = \|H( w_{c,\angle90^\circ})-H(w_{\angle0^\circ})\|^2,
	\label{eq:L_f}
\end{equation}
where $w_{c,\angle90^\circ}$ is the predicted clean image, $H()$ is the operation of summing each row of an image along the width direction. For an $H\times W$ image, this operation transforms it into a vector of size $H\times1$.

The overall loss function can be formulated as:
\begin{equation}
	\mathcal{L}_{SvDiff}= \lambda_{noise}\mathcal{L}_{noise} + \lambda_{img}\mathcal{L}_{img} + \lambda_{sp}\mathcal{L}_{sp} + \lambda_{vel}\mathcal{L}_{vel}.
	\label{eq:L_all}
\end{equation}
By gradient steps on these losses, $\text{SvDiff}$ generates physically accurate and visually realistic side-view predictions. Our multi-frame training strategy explicitly encourages temporal consistency, ensuring coherent and stable smoke motion.

\subsection{Progressive Novel View Refinement}
\label{subsec:refinement}

Based on 2D images from various views, we can train a density generator $\mathcal{G}_{\rho}$ to estimate a 3D density field of smoke as:
\begin{equation}
	\rho_r^t=\mathcal G_{\rho}(I^t),\quad I^t = w^t_{\angle0^\circ} \oplus w^t_{p,\angle90^\circ} \oplus \cdots.
\end{equation}
Here $\mathcal{G}_{\rho}$ adopts the UNet3+ architecture~\cite{huang2020unet} and extends the 2D convolutions in UNet3+ to 3D convolutions. 
Please refer to Appendix for more details.
Since estimating density along ray direction from 2D images is difficult, we design the following loss for $\mathcal{G}_{\rho}$:
\begin{equation}
	\begin{split}
		\mathcal {L}_{\mathcal G_\rho}=&
		\lambda_{\rho} \| \rho_r^t -{\rho}^t\|^2+ \lambda_{in}  \sum_{\alpha \in \mathbb A} \| \mathcal R(\rho_r^t,\alpha) -\mathcal R({\rho}^t,\alpha)\|^2 \\&+\lambda_{un} \sum_{\alpha \notin \mathbb A}\| \mathcal R(\rho_r^t,\alpha) -\mathcal R({\rho}^t,\alpha)\|^2,
		\label{eq:L_grho}
	\end{split}
\end{equation}
where $\rho$ denotes the ground truth density, $\mathbb{A}$ denotes the set of input view angles (e.g., $\angle0^\circ, \angle90^\circ$), $\mathcal{R}(\rho,\alpha)$ is the differentiable rendering operator that renders density field $\rho$ at the viewing angle $\alpha$.
The second and third terms correspond to images from input and unknown viewpoints, respectively.
For the ScalarFlow dataset, we set $\lambda_{\rho}$ to zero and use the reconstructed results from~\cite{eckert2019scalarflow} as ${\rho}$ for rendering.
In our pipeline, when the number of input images is less than 16, we call it coarse-grained density generator $\mathcal G^c_{\rho}$; when the number of input images equals to 16, we call it fine-grained density generator $\mathcal G^f_{\rho}$. 

After generating side-view $ w^t_{p,\angle90^\circ}$ from $w^t_{\angle0^\circ}$ with $\text{SvDiff}$, we employ $\mathcal G^c_{\rho}$ to produce rough density field $\rho^i_{r,c}$. Although $\mathcal G^c_{\rho}$ is trained using the rendered image loss $\| \mathcal R(\rho^t,\alpha) -\mathcal R({\rho}^t_{r,c},\alpha)\|^2$ to learn the smoke shape in novel views, in the absence of enough views, $\rho^t_{r,c}$ still exhibits blurriness in novel views.

To enhance details and reduce blurriness in $\rho_{r,c}$, we introduce novel view refinement module $\text{NvRef}$ based on UNet3+:
\begin{align}
    \centering
	\begin{split}
        res_\alpha^t &=  \text{NvRef}\big( w^t_{r,\angle\alpha-\beta} \oplus  w^t_{r,\angle\alpha+\beta} \oplus  w^t_{r,\angle \alpha} \\
        &\qquad\qquad  \oplus \downarrow w^{t-1}_{f,\angle \alpha} \oplus \downarrow  w^{t-2}_{f,\angle \alpha}\big), \\ 
         {w}^t_{f,\angle \alpha} &=  res_\alpha^t + w^t_{r,\angle \alpha},
	\end{split}
\end{align}
where $\alpha$ is the target angle to be refined, $\beta$ is the angular offset relative to $\alpha$, $\downarrow$ is 2x downsampling operation, and $res$ is the residual error. 

\text{NvRef} is designed to maintain the spatial distribution consistency and perceptual similarity between ground truth and refined novel images, whose overall loss function is formulated as:
\begin{small}
\begin{align}
	\begin{split}
	\mathcal{L}_{NvRef}& = \lambda_{mse}\|{w}^t_{f,\angle \alpha} - {w}^t_{\angle \alpha} \|^2 +  \lambda_{l1}\|{w}^t_{f,\angle \alpha} - {w}^t_{\angle \alpha} \| \\
	& + \lambda_{res}\|Mean(res^t_\alpha)\|^2 + \lambda_{sp}\|H({w}^t_{f,\angle \alpha})-H({w}^t_{\angle \alpha})\|^2 \\
	&+  \lambda_{psnr}\|PSNR({w}^t_{f,\angle \alpha})-PSNR({w}^t_{\angle \alpha})\|^2,
	\end{split}
\end{align}
\end{small}
where the first three terms penalize $L2$, $L1$ and residual error, the fourth is spatial constraint similar to SvDiff, and the last computes the peak signal-to-noise ratio (PSNR) discrepancy.

\begin{figure}[htb] 
    \centering
	\includegraphics[width=0.4\textwidth]{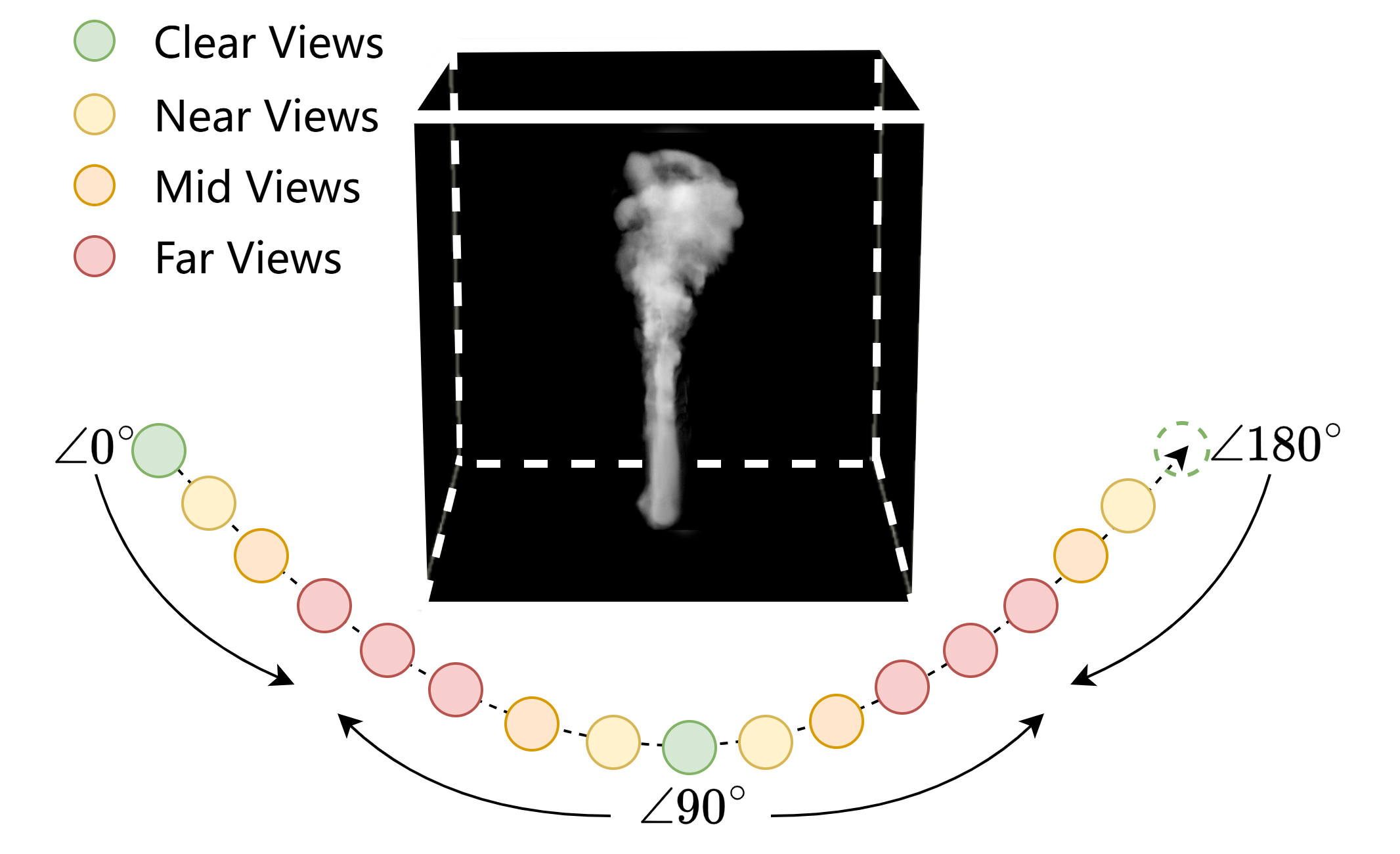}
	\caption{The progressive scheme for novel view refinement begins with clear views and incrementally rotates the camera to render and refine novel-view images from near, mid, and far views.}
	\label{fig:progressive_refinemnt}
\end{figure}

Subsequently, we iteratively invoke $\mathcal{G}\rho$ and \text{NvRef} to rotate the camera along the horizontal plane, progressively rendering and refining additional novel view images. In our experiments, we set the maximum number of views to 16 to achieve a balance between computational efficiency and reconstruction quality. Since rendered images from adjacent views tend to exhibit similar shapes and reduced blurriness, we further categorize these 16 views into four types, namely clear, near, mid, and far views, based on their relative positions to the front and side views, as illustrated in Fig.~\ref{fig:progressive_refinemnt}. 

During multi-stage refinement process, we sequentially render images at near, mid, and far views from the density field reconstructed in the previous stage, and refine these images using \text{NvRef}. The refined images, together with the blurred images from the remaining views, are then used to reconstruct the density field for the next stage of refinement. By iteratively combining coarse 3D density estimation with targeted refinement of novel view images, our progressive novel view refinement strategy gradually expands the set of reliable views. Finally, we leverage multi-view information to jointly reconstruct the density, velocity, and inflow of the input smoke phenomena. See supplementary for details.

\section{Evaluations and Ablation Study}
\label{sec:experiments}

\subsection{Evaluation}
\label{subsec:Evaluation}

\paragraph{Evaluation on ScalarFlow.}
To validate the applicability of our method in real-world scenarios, we conducted evaluations on the ScalarFlow dataset~\cite{eckert2019scalarflow}. This dataset captures real-world smoke images using five cameras uniformly distributed along a $120^\circ$ arc and provides 3D density and velocity fields. However, these 3D data cannot be directly used for quantitative comparison, so our subsequent evaluations are based solely on images.

\begin{figure}[htb]
	\centering
	\includegraphics[width=1.0\linewidth]{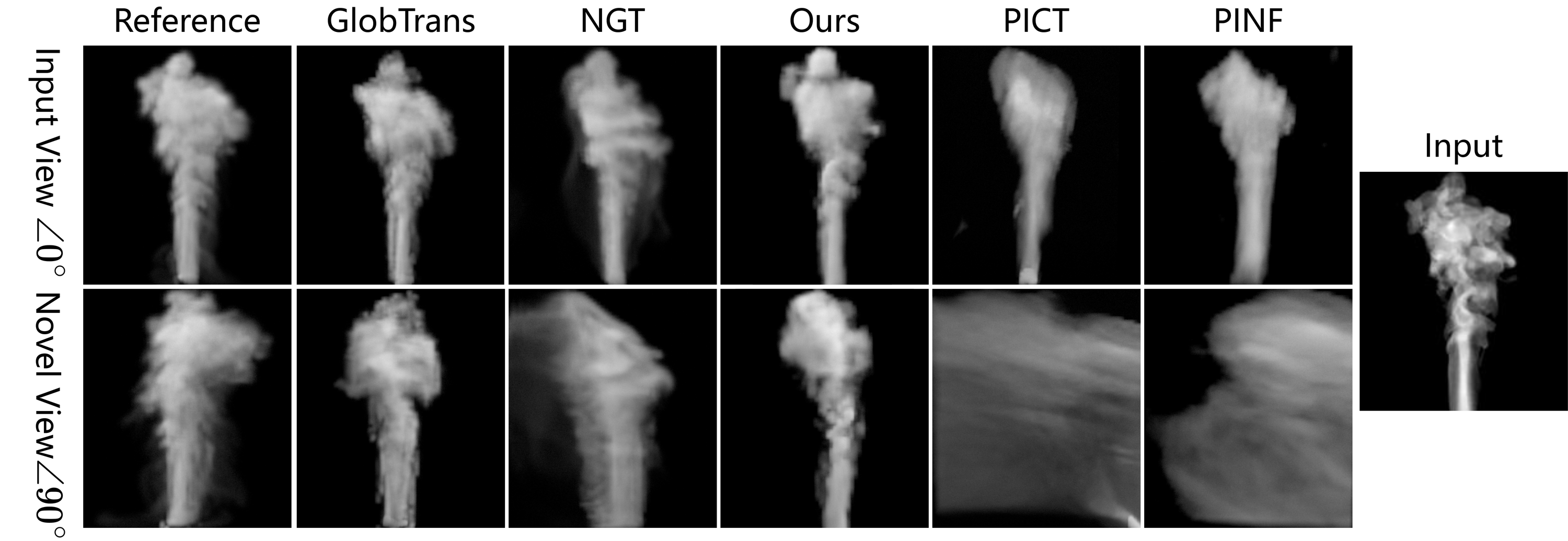}
	\caption{Qualitative comparison based on different methods on ScalarFlow. Our method matches the appearance pattern of the input image at the front view, and produces a reasonable shape in the side view.}
	\label{fig:compare_SF}
\end{figure}

\begin{figure}[htb]
	\centering
	\includegraphics[width=0.9\linewidth]{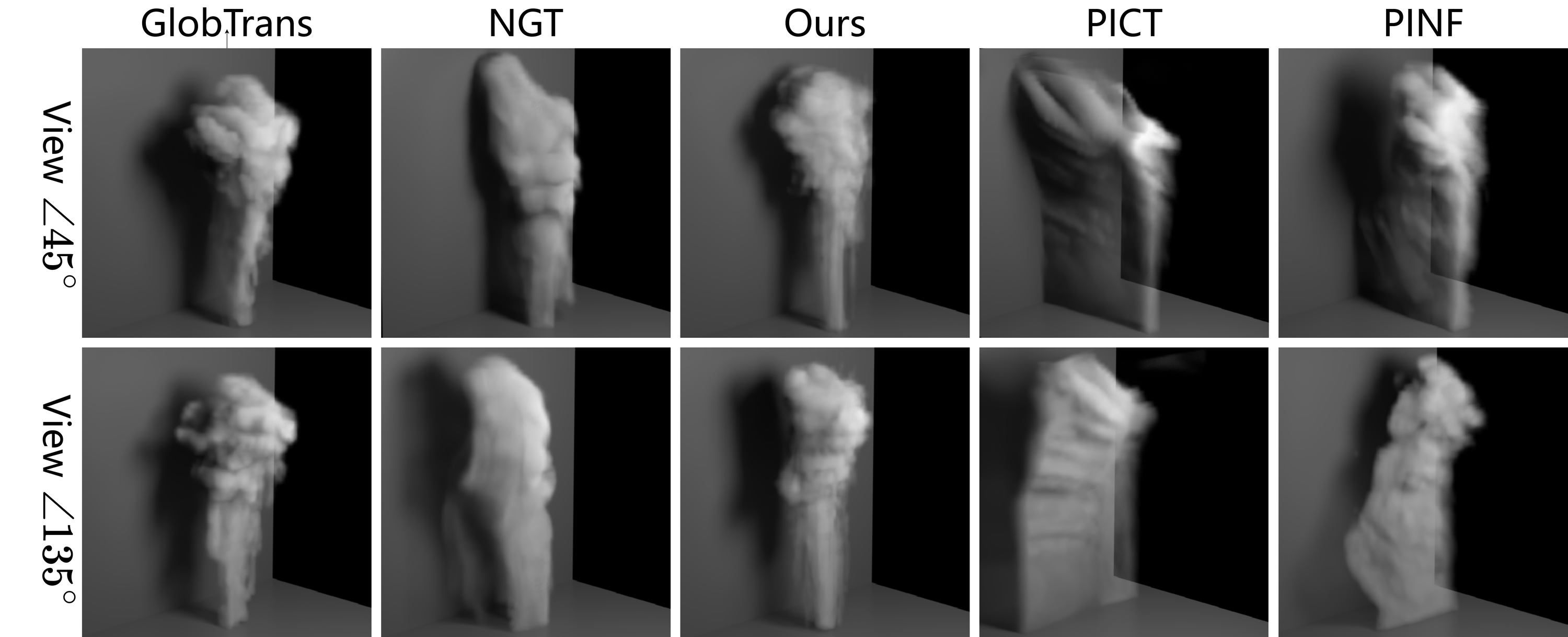}
	\caption{Qualitative comparison on ScalarFlow.}
	\label{fig:compare_SF_45/135}
\end{figure}

In our experiments, we used one of the pre-processed images from the five viewpoints in the ScalarFlow as input to reconstruct smoke density fields at a resolution of \(64 \times 112 \times 64\). For comparison, we interpolated the density fields reconstructed by all methods to the same resolution of $64^3$ and rendered images at the input front view ($\angle 0^\circ$) and side view ($\angle 90^\circ$) using Houdini. We conducted qualitative comparisons with state-of-the-art methods, as shown in Figs.~\ref{fig:compare_SF} and~\ref{fig:compare_SF_45/135}. Due to limited single-view input, PICT and PINF exhibit varying degrees of blurring in the depth direction, even affecting the reconstruction quality at the front view. In contrast, GlobTrans achieves the best perceptual quality (as documented in Table~\ref{tb:SF_test}) at the side view and performs well across multiple novel views, at the expense of heavy computational cost. The results of NGT match well with inputs through differentiable rendering and adversarial learning techniques, achieving the lowest root mean square error at novel views. However, it introduces artifacts in certain views ($90^\circ$ in Fig.~\ref{fig:compare_SF}) and presents overly smooth smoke at some angles ($135^\circ$ in Fig.~\ref{fig:compare_SF_45/135}).

These results indicate the difficulty of balancing reconstruction quality and computational efficiency from single-view input. Our method matches input images well while maintaining reasonable smoke appearance and rich details in novel views at minimal cost.  From a perceptual quality perspective, our method performs excellently, second only to GlobTrans. However, as shown in Table~\ref{tb:SF_test}, mean squared error cannot comprehensively measure novel view quality—PICT and PINF exhibit unreasonable appearance yet achieve similar MSE to our method.

\begin{table}[!ht]
	\centering
	\caption{Quantitative comparison on ScalarFlow.}
	\resizebox{0.5\textwidth}{!}
	{
        \begin{tabular}{c|c c c c|c c|c}  
            \hline  
            Algorithm & \makecell{Input \\ RMSE}$\downarrow$ & SSIM $\uparrow$& PSNR$\uparrow$ & LPIPS$\downarrow$ & \makecell{Side \\ RMSE}$\downarrow$ & STYLE$\downarrow$ &\makecell{Time for \\ 120 Steps}\\  
            \hline
            GlobTrans & \pmb{0.0101} & \pmb{0.9975} & \pmb{40.1560} & \pmb{0.0054} & \pmb{0.0352} & \underline{0.2167} & $>$30h%
            \\
            NGT & 0.0289 & 0.9539 & 31.0727 & 0.0655 & \underline{0.0544} & 0.2499 & 5mins\\
            PICT  &0.0315 & 0.9252 & 30.5447 & 0.1332 & 0.0743 & 0.7259 & /\\
            PINF &0.0872 & 0.8715 & 21.3005 & 0.1020 & 0.1101 & 0.6335 &/\\
            Ours & \underline{0.0127} & \underline{0.9868} & \underline{38.0790} & \underline{0.0223} & 0.0853 & \pmb{0.2071} & 15mins \\
            \hline
            \end{tabular}
	}
	\label{tb:SF_test}
\end{table}

\begin{figure}[htb]
	\centering
	\includegraphics[width=0.85\linewidth]{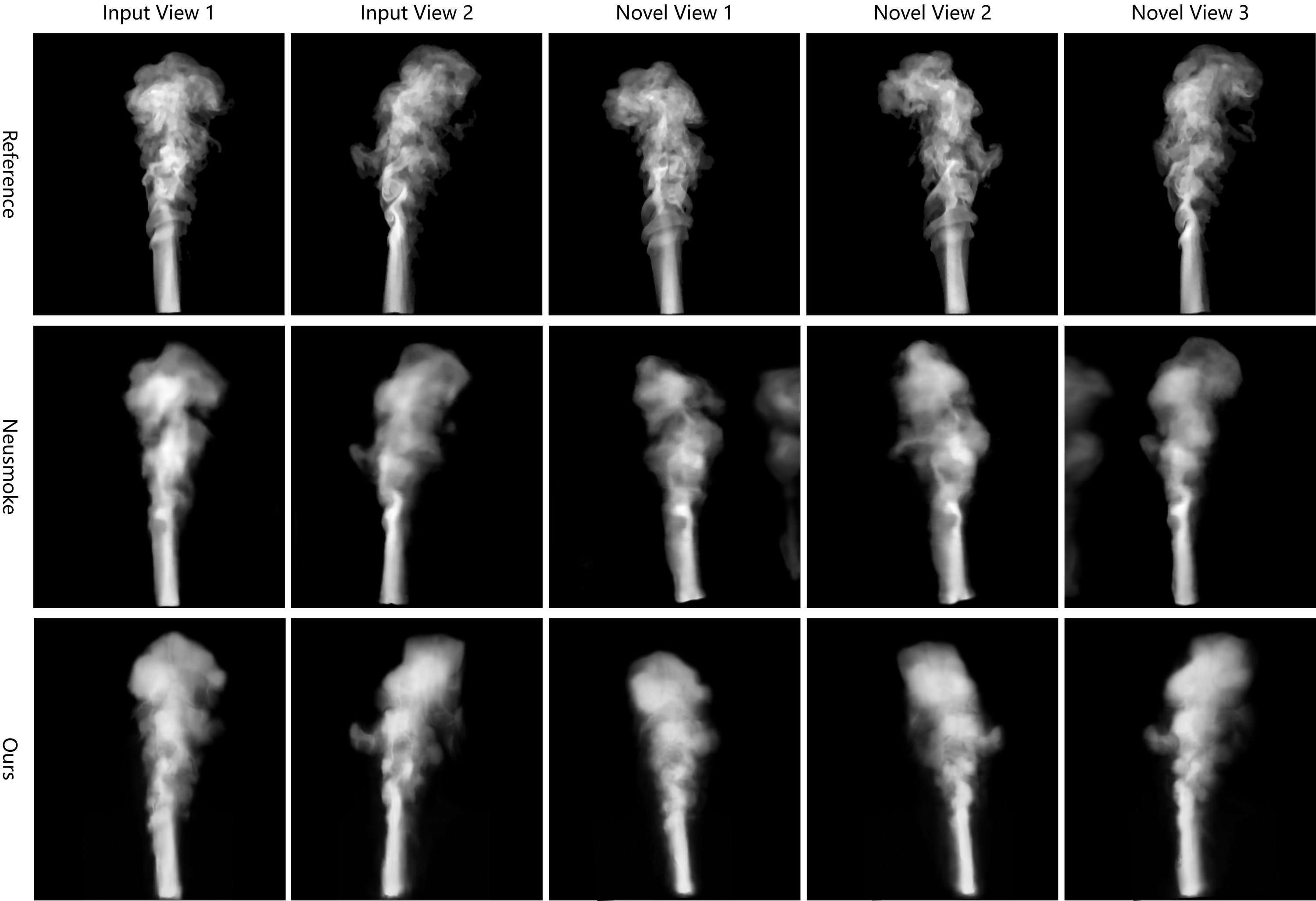}
	\caption{Comparison with NeuSmoke on ScalarFlow.}
	\label{fig:compare_Neusmoke}
\end{figure}

Tables~\ref{tb:fluidnexus} and~\ref{tb:neusmoke} compare our method with FluidNexus~\cite{gao2025fluidnexus} and NeuSmoke~\cite{qiu2024neusmoke}. Our method significantly outperforms both approaches on input view reconstruction across all metrics. Compared to NeuSmoke, we achieve substantial improvements on novel views, demonstrating that our progressive refinement strategy, which explicitly synthesizes side views, effectively alleviates single-view ambiguity better than implicit neural rendering from sparse views. For FluidNexus, while our novel view performance is slightly lower (as its multi-view diffusion inherently maintains cross-view consistency), we achieve superior input quality through progressive side-view refinement and avoid sensitivity to post-processing threshold selection. Our novel view refinement module further enhances quality through multi-view consistency constraints, producing accurate reconstructions without requiring hyperparameter tuning, demonstrating superior robustness. The qualitative comparison is shown in Figs.~\ref{fig:compare_fluidnexus} and~\ref{fig:compare_Neusmoke}.

\begin{table}[!ht]
\caption{Comparison with FluidNexus (various post-processing thresholds) on ScalarFlow. Averaged over five scenes, novel views from four non-frontal cameras.}
\resizebox{0.5\textwidth}{!}{
\begin{tabular}{c|cccc|cccc}
\hline
 Algorithm & \makecell{Input \\ RMSE}$\downarrow$ & SSIM $\uparrow$& PSNR$\uparrow$ & LPIPS$\downarrow$ & \makecell{Novel \\ RMSE}$\downarrow$ & SSIM $\uparrow$& PSNR$\uparrow$ & LPIPS$\downarrow$ \\
\hline
FN w/o th & 0.0473 & 0.7924 & 26.6722 & 0.2192 & 0.0807 & 0.1651 & 21.9411 & 0.1881 \\
FN th=0.05 & \underline{0.0303} & 0.8858 & \underline{30.8166} & 0.1912 & 0.0702 & 0.3187 & 23.2492 & 0.2665 \\
FN th=0.1 & 0.0388 & \underline{0.9159} & 30.7635 & \underline{0.1217} & \pmb{0.0565} & \pmb{0.8419} & \pmb{25.3569} & \underline{0.1575} \\
FN th=0.15 & 0.0361 & 0.8968 & 29.3974 & 0.1402 & \underline{0.0582} & 0.8435 & \underline{25.1001} & \pmb{0.1573} \\
FN th=0.2 & 0.0428 & 0.8757 & 27.8309 & 0.1628 & 0.0598 & \pmb{0.8419} & 23.9521 & 0.1669 \\
ours & \pmb{0.0172} & \pmb{0.9764} & \pmb{35.5504} & \pmb{0.0586} & 0.0690 & 0.7871 & 23.4393 & 0.1829 \\
\hline
\end{tabular}
}
\label{tb:fluidnexus}
\end{table}

\begin{table}[!ht]
	\tiny
   \caption{Comparison with Neusmoke on ScalarFlow. Front and side views as input, three novel views for evaluation (averaged).}
    \centering
\resizebox{0.5\textwidth}{!}{
	
   \begin{tabular}{c|cccc}
\hline
 Algorithm & RMSE$\downarrow$ & SSIM $\uparrow$& PSNR$\uparrow$ & LPIPS$\downarrow$ \\
\hline
NeuSmoke & 0.0514&0.8750&26.5031&0.1131\\
Ours & \pmb{0.0331} & \pmb{0.9038} & 
\pmb{30.0384} & \pmb{0.0991}\\
\hline
\end{tabular}}
\label{tb:neusmoke}
\end{table}

\begin{figure}[htb]
	\centering
	\includegraphics[width=0.85\linewidth]{figures/compare_fluidnexus.png}
	\caption{Comparison with FluidNexus on ScalarFlow.}
	\label{fig:compare_fluidnexus}
\end{figure}

\paragraph{Evaluation on Synthetic Data.}
We evaluated our method on a synthetic smoke dataset generated with the rendering operator~\cite{franz2021global}. The synthetic dataset provides precise 3D physical fields and smooth motion compared to real-world scenes. Table~\ref{tb:syn_test} shows performance comparison with baseline methods using image metrics.

\begin{figure}[htb]
	\centering
	\includegraphics[width=0.9\linewidth]{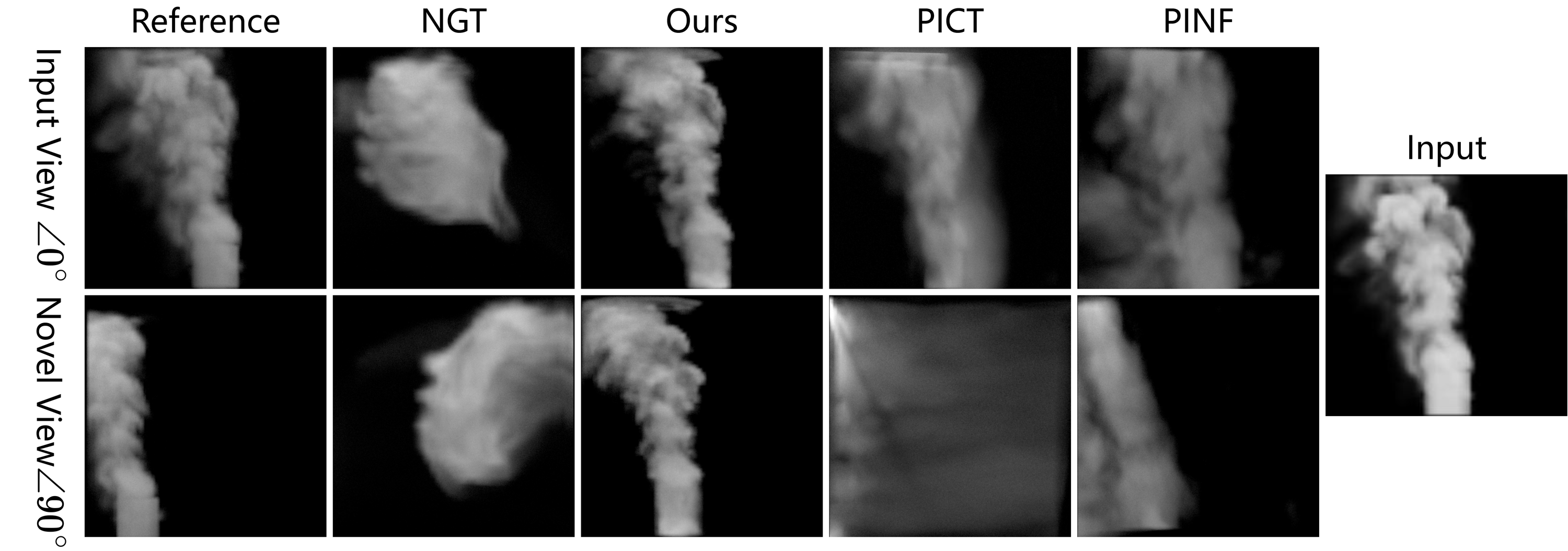}
	\caption{Qualitative comparison on the synthetic dataset.}
	\label{fig:compare_syn}
\end{figure}

Fig.~\ref{fig:compare_syn} shows qualitative comparison with state-of-the-art methods. Similar to ScalarFlow results, PICT and PINF exhibit blurriness in side views. Additionally, NGT's inaccurate inflow estimation causes reconstructed density to gradually deviate from input over time. See Sec. E in supplementary for more complex phenomena.

\begin{table}[!ht]
	\centering
	\caption{Quantitative comparison on the synthetic dataset.}
	\resizebox{0.5\textwidth}{!}{
		\begin{tabular}{c|c c c c|c c} 
			\hline  
			Algorithm & \makecell{Input \\ RMSE}$\downarrow$ & SSIM $\uparrow$& PSNR$\uparrow$ & LPIPS$\downarrow$ & \makecell{Side \\ RMSE}$\downarrow$ & STYLE$\downarrow$ \\  
			\hline
			NGT & 0.1844 & \underline{0.7754} & 15.6521 & 0.2227 & \underline{0.2714} & {1.2242} \\
			PICT  & \underline {0.1625} & 0.7608 & \underline{16.2969} & \underline{0.2153} & 0.2913 & 1.5585\\
			PINF &0.2286 & 0.6293 & 13.2970 & 0.2259 & \pmb {0.2468} & \underline {1.1321}\\
            Ours & \pmb{0.0395} & \pmb{0.9645} & \pmb{28.1332} & \pmb{0.0293} & 0.3821 & \pmb{1.0790} \\
			\hline
		\end{tabular}
	}
	\label{tb:syn_test}
\end{table}

\begin{figure}[htb] 
	\centering
	\includegraphics[width=1\linewidth]{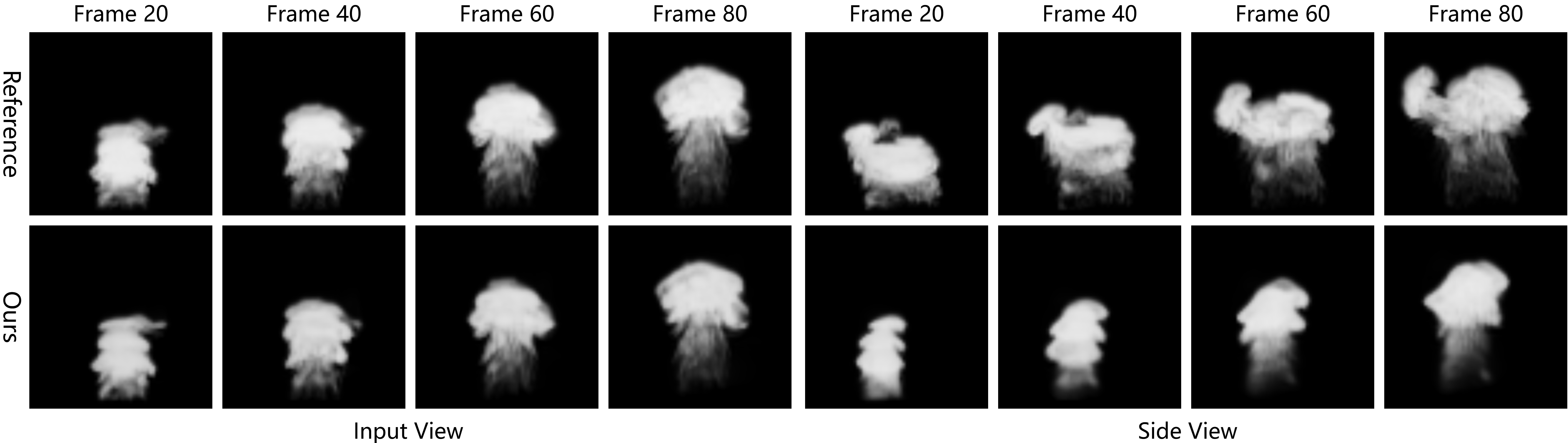}
	\caption{Reconstruction results for a bunny-shaped smoke scenario without inflow.}
	\label{fig:no_source}
\end{figure}

\begin{figure}[htb]
	\centering
	\includegraphics[width=1\linewidth]{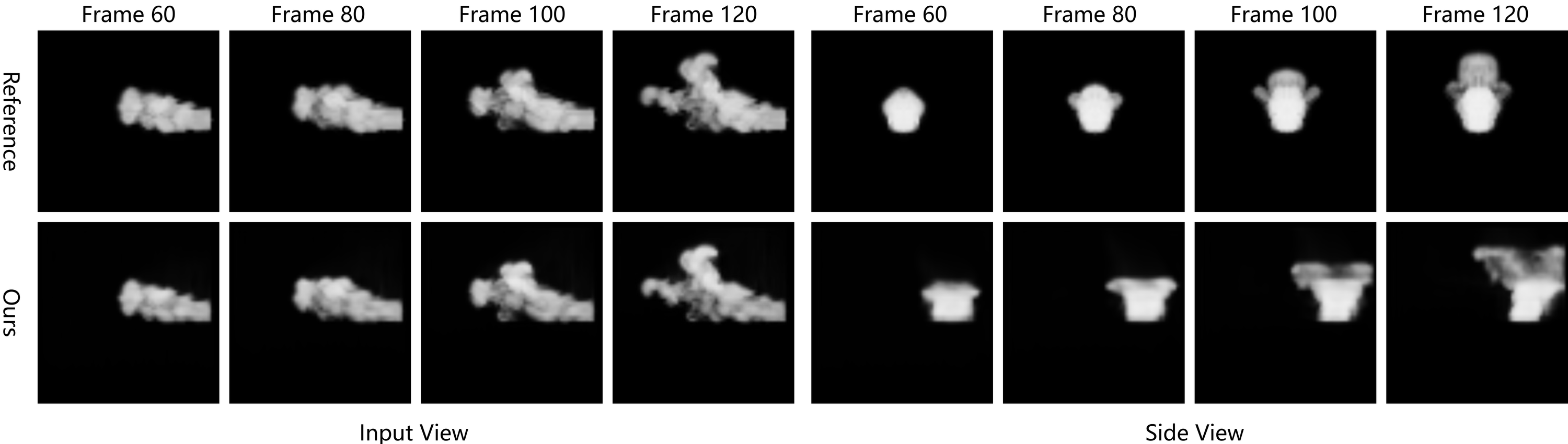}
	\caption{Reconstruction result for a horizontal plume scenario.}
	\label{fig:horizontal_smoke}
\end{figure}

\paragraph{Generalization Performance. }
To evaluate generalization, we apply our method to smoke without inflow and horizontal plume (Figs.~\ref{fig:no_source} and~\ref{fig:horizontal_smoke}), unseen in training. Results show effectiveness on these previously unseen scenarios.

\subsection{Ablation Study}
\label{subsec:ablation}

\paragraph{Ablation on Side-view Synthesizer.}
To evaluate physical priors in SvDiff, we remove noise threshold, velocity loss, gradient loss, and 3D reconstruction ("w/o threshold", "w/o vel", "w/o grad", "w/o reconstruction"). Table~\ref{tb:ablation_svdiff} shows removing these constraints degrades performance. Note that velocity-based temporal correction slightly reduces input view LPIPS.

\begin{table}[!ht]
\centering
\caption{Ablation studies on SvDiff.}
\resizebox{0.5\textwidth}{!}{%
\begin{tabular}{c|c c c c|c c}  
\hline  
Algorithm & \makecell{Input \\ RMSE}$\downarrow$ & SSIM $\uparrow$& PSNR$\uparrow$ & LPIPS$\downarrow$ & \makecell{Side \\ RMSE}$\downarrow$ & STYLE$\downarrow$ \\  
\hline
w/o threshold & \underline{0.0089} & \underline{0.9946} & 41.8412 & 0.0096 & \underline{0.0990} & 0.2139 
\\
w/o vel & 0.0100 & 0.9929 & 41.6814 & \underline{0.0069} & 0.1032 & 0.2074\\
w/o grad & 0.0091 & 0.9940 & \underline{42.0804} & \pmb{0.0061} & 0.1025 & \underline{0.2025} \\
w/o divergence & 0.0136 & 0.9886 & 40.9043 & 0.0114 & 0.1816 & 0.4831 \\
w/o reconstruction &0.0106 & 0.9934 & 41.4763 & 0.0077 & 0.1025 & 0.3118  \\
Ours & \pmb{0.0062} & \pmb{0.9955} & \pmb{44.5518} & {0.0075} & \pmb{0.0899} & \pmb{0.1892} \\
\hline
\end{tabular}
}
\label{tb:ablation_svdiff}
\end{table}

\begin{figure*}[htb]
	\centering
	\includegraphics[width=0.8\linewidth]{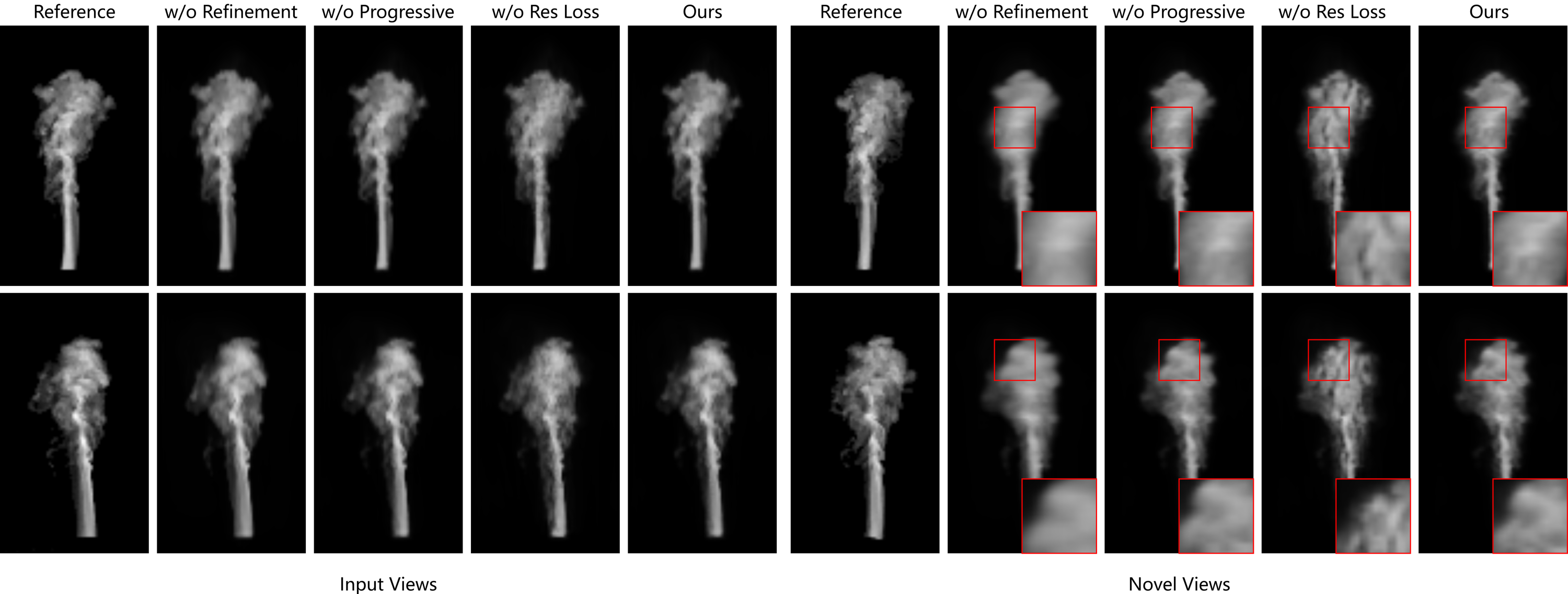}
	\caption{Ablation on novel view refinement. From top to bottom: reference, results without refinement, without progressive refinement, without res loss and with NvRef. Red boxes show close-ups.}
	\label{fig:ablation_refinemnt_visual}
\end{figure*}

\begin{figure}[htb]
	\centering
	\includegraphics[width=0.8\linewidth]{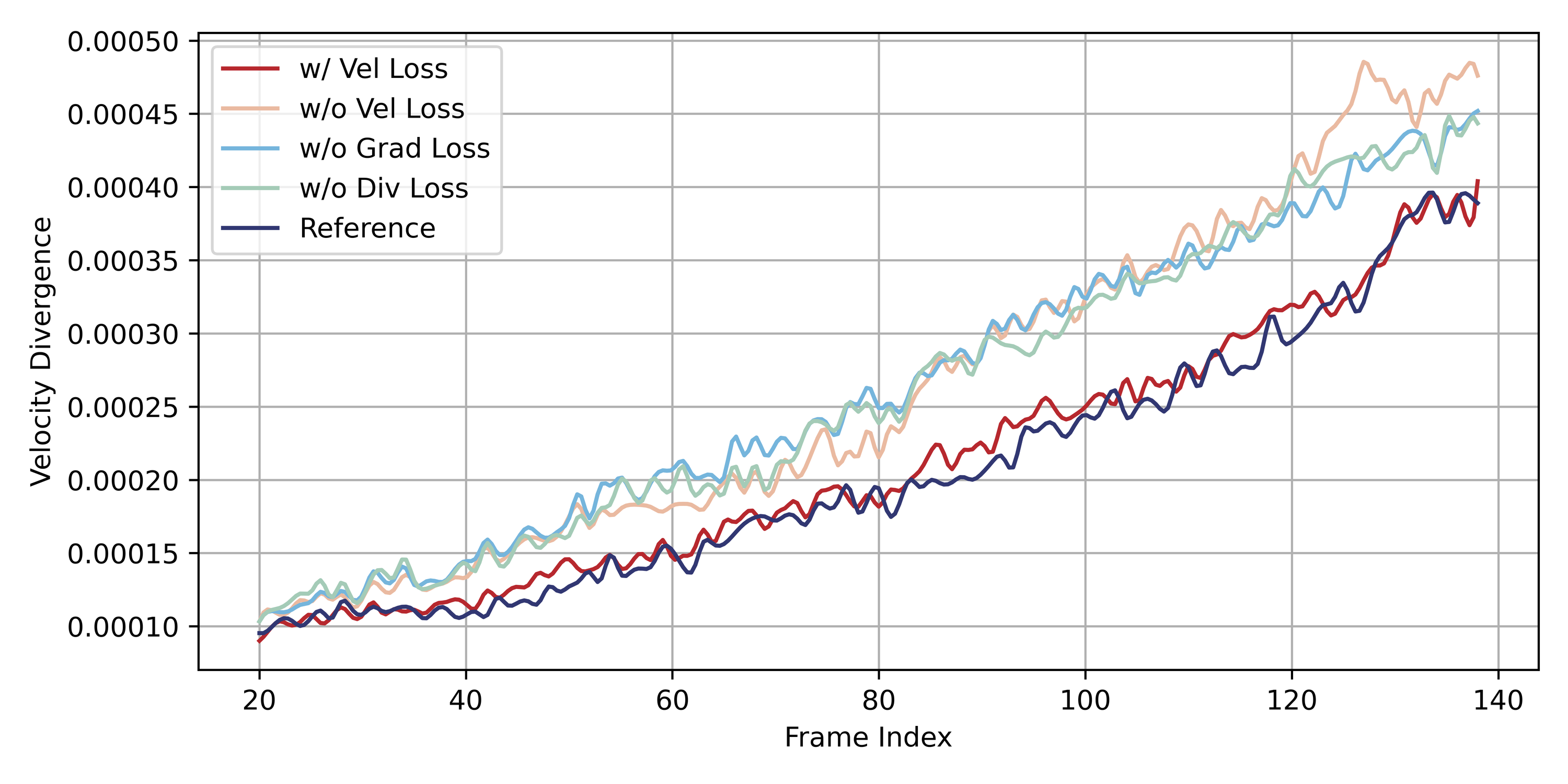}
	\caption{Comparison of the divergence of reconstructed velocity fields by SvDiff with different loss functions at various time steps.}
	\label{fig:ablation_svdiff_veldiv}
\end{figure}

\begin{wrapfigure}{r}{1.0in}
	\centering
	\includegraphics[width=1.0in]{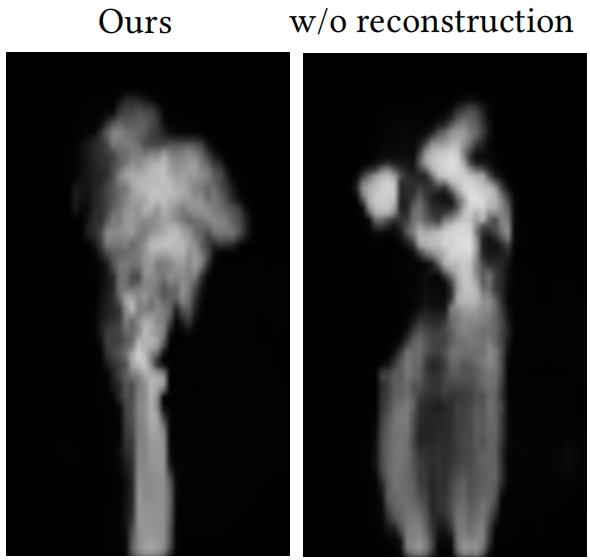}
	\caption{Ablation on rendered density input for side-view synthesis. }
	\label{fig:ablation_rendered_image}
\end{wrapfigure}

Fig.~\ref{fig:ablation_svdiff_veldiv} visualizes the divergence of reconstructed velocity fields to demonstrate the velocity term's impact. Incorporating velocity loss produces smoother and more stable smoke dynamics, preventing artifact flickering. To evaluate visual priors, we ablated rendered density images as SvDiff input. Fig.~\ref{fig:ablation_rendered_image} shows omitting these images causes noticeable errors in long-term synthesis.

\paragraph{Ablation on Novel View Refinement.}

\begin{figure}[htb]
	\centering
	\includegraphics[width=0.85\linewidth]{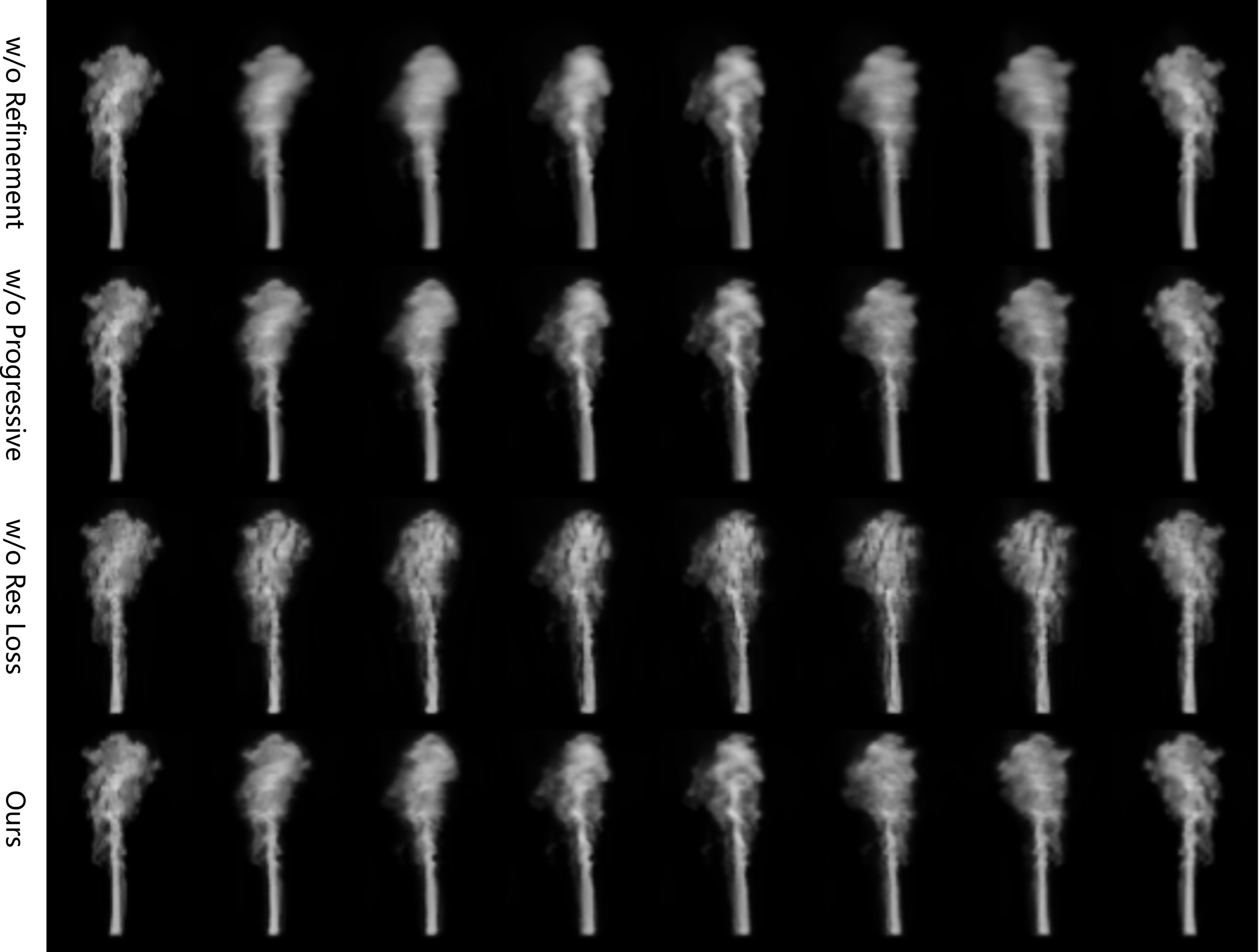}
	\caption{Refined results across novel views. Each row shows renderings uniformly distributed from \(\angle0^\circ\) to \(\angle175^\circ\).}
	\label{fig:ablation_refinemnt_moreviews}
\end{figure}

To assess the impact of novel view refinement, we performed ablation studies by (1) removing the entire refinement process, (2) replacing the multi-stage progressive refinement with a single-pass refinement for all novel views, and (3) remove residual loss. These variants are denoted as "w/o Refinement", "w/o Progressive," "w/o Res Loss", respectively. The quantitative and qualitative results are presented in Table~\ref{tb:ablation_refinement}, Figs.~\ref{fig:ablation_refinemnt_visual} and~\ref{fig:ablation_refinemnt_moreviews}. Our progressive refinement approach achieves richer visual details and appearance consistency.

\begin{table}[!ht]
\centering
\caption{Ablation on novel view refinement. Views 0 (front) and 3 (side) as input, remaining views for evaluation.}
\resizebox{0.5\textwidth}{!}{
\begin{tabular}{c|c c c c}  
\hline  
Algorithm & MSE $\downarrow$ & SSIM $\uparrow$ & PSNR $\uparrow$ & LPIPS $\downarrow$\\  
\hline
w/o Refinement & 0.0196 & 0.7454 & 18.7490 & 0.1808  \\
w/o Progressive & {0.0192} & \pmb{0.7559}& \underline{18.7902} &\pmb{0.1704}   \\
w/o Res Loss & \pmb{0.0168} & 0.7126 & 18.5066 & 0.1789   \\
Ours & \underline{0.0190} & \underline{0.7559} & \pmb{18.7978} & \underline{0.1757} \\
\hline
\end{tabular} 
}
\label{tb:ablation_refinement}
\end{table}

\begin{figure}[htb]
	\centering
	\includegraphics[width=0.9\linewidth]{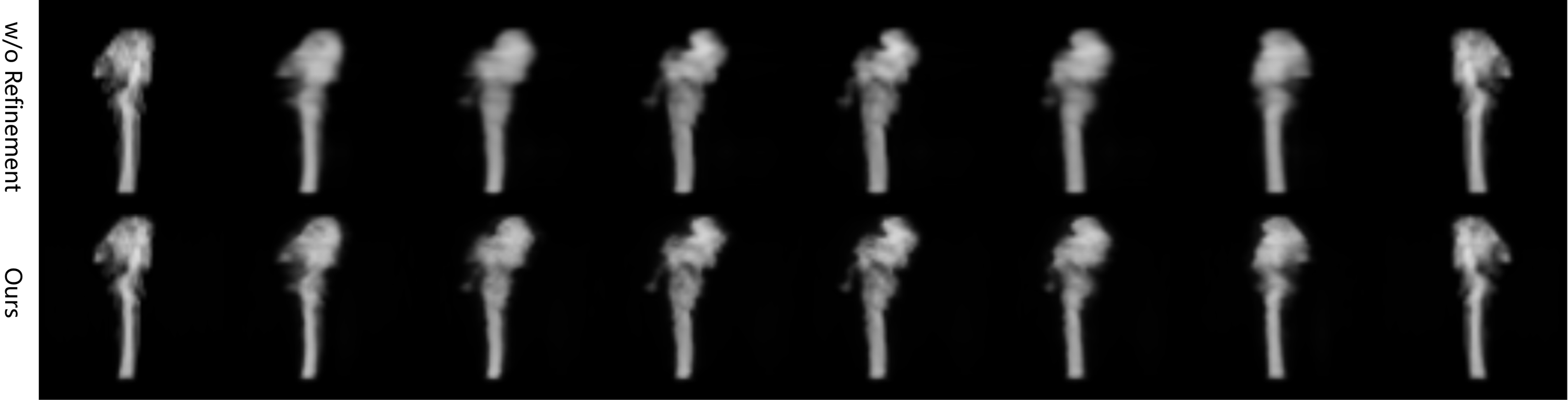}
	\caption{Ablation study of the refinement model. Each row shows renderings uniformly distributed from $\angle0^\circ$ to $\angle175^\circ$.}
	\label{fig:ablation_refinement_usePredictor}
\end{figure}

\paragraph{Ablation on Key Components.}
To evaluate key components, we conduct two ablation studies: (1) removing novel view refinement, and (2) replacing our side-view synthesizer with NGT~\cite{franz2023learning}. 
Fig.~\ref{fig:ablation_refinement_usePredictor} shows novel views before and after refinement, demonstrating that refinement produces richer details and reduces blurriness.

\section{Conclusion and Future Work}
\label{sec:conclusion}

We present a framework for 3D smoke reconstruction from single-view input by integrating physical priors and spatiotemporal constraints. Our approach overcomes single-view ambiguity through a diffusion-based side-view synthesizer and novel view refinement module, providing rich multi-view information for density and velocity reconstruction. Experiments on synthetic and real-world datasets demonstrate superior balance between quality and efficiency. Our framework maintains accurate input matching while preserving reasonable smoke appearance and rich details in novel views. Future work could address more complex fluids, vertical multi-view fusion, and higher-order physical constraints.

\section*{Acknowledgements}
This research was supported by Zhejiang Provincial Natural Science Foundation of China under Grant No.ZCLQN26F0204, the Open Project Program of State Key Laboratory of Virtual Reality Technology and Systems, Beihang University (No.VRLAB2025C05), National Natural Science Foundation of China (No.U25A20444, No.62372325, No.62402255, No.62502344), Natural Science Foundation of Tianjin Municipality (No.23JCZDJC00280), Shandong Provincial Natural Science Foundation (No.ZR2024QF020), Shandong Province National Talents Supporting Program (No.2023GJJLJRC-070), Young Talent of Lifting engineering for Science and Technology in Shandong (No.SDAST2024QTB001), Shandong Project towards the Integration of Education and Industry (No.2024ZDZX11). 

{
    \small
    \bibliographystyle{ieeenat_fullname}
    \bibliography{main}
}



\end{document}


\maketitle  

\renewcommand{\thefootnote}{\fnsymbol{footnote}}
\footnotetext[1]{Corresponding Author. 	$^\dag$Equal contributions. }
\renewcommand{\thefootnote}{\arabic{footnote}}

\section*{A. Overview}
In this supplementary material, we provide additional background, detailed descriptions of the technical approach, implementation specifics, evaluation results, and ablation studies. We also discuss the limitations of our work and outline potential directions for future research.

\section*{B. Preliminary}
\label{sec:preliminary}

\paragraph{Navier-Stokes Equation.}
Generally, fluid motion is governed by the well-known incompressible Navier-Stokes equations:
\begin{align}
	\frac{\partial \mathbf{u}}{\partial t}+(\mathbf{u} \cdot \nabla)\mathbf{u}&=-\frac{\nabla p}{\rho} + \nu \nabla^2 \mathbf{u}+\mathbf{f},
	\label{eq:ns}
	\\\nabla \cdot \mathbf{u}&=0,
	\label{eq:div}
\end{align}
where $\mathbf{u}$ is the velocity, $\rho$ is the density, $p$ is the pressure, $\mathbf{f}$ is the external force, and $\nu$ is the viscosity coefficient, which is usually set to zero for smoke phenomena. Eq.~\ref{eq:ns} is the momentum equation, which describes the time rate of velocity change, while Eq.~\ref{eq:div} is the mass conservation equation to preserve the incompressibility. 
To formalize, density evolution follows the transport equation:
\begin{equation}
	\frac{\partial \rho}{\partial t}+\mathbf{u}\cdot \nabla \rho = 0.
	\label{eq:transport}
\end{equation}

\paragraph{Diffusion Models.}
Diffusion probabilistic models (DDPM) consist of two processes: a forward diffusion process and a reverse inference process. During the training stage, given a data point \(x_0 \sim q(x)\) sampled from the real data distribution, the forward process adds Gaussian noise to the sample \(x_0\) over \(S\) time steps, constructing a Markov chain diffusion process:
\begin{align}
	q(x_s|x_{s-1})&=\mathcal{N}(x_s;\sqrt{1-\beta_s}x_{s-1},\beta_s\mathit{I}),\\
	q(x_{1:S}|x_0)&=\prod^S_{s=1}q(x_s|x_{s-1}),
\end{align}
where \(\mathcal{N}\) denotes a Gaussian distribution, \(\beta_s\) denotes a fixed or learnable variance schedule parameter that controls the noise intensity added at each step, \(x_s\) denotes the noisy image at time step \(s\) (selected from the total steps \(S\)), which can be expressed as:
\begin{equation}
	x_s = \sqrt{\bar \alpha_s} x_0 + \sqrt{1-\bar \alpha_s} \epsilon,
	\label{noiseImage}
\end{equation}
where $\alpha_s=1-\beta_s$, $\bar \alpha_s:= \prod^{s}_{s=1}\alpha_s$, and $\epsilon \sim \mathcal{N}(0, \mathit{I})$. The model is trained to minimize the following loss function:
\begin{equation}
	\|\epsilon-\epsilon_\theta (x_s,s) \|^2.
	\label{eq:diffusion_loss}
\end{equation}
During the generation stage, the diffusion model samples a Gaussian random noise \(x_S \sim \mathcal{N}(0,I)\), and utilizes the predefined variance \(\sigma_s\) and random noise \(\epsilon_s\) to gradually denoise it to until \(x_0\). This process is formulated as:
\begin{equation}
	\begin{split}
		x_{s-1}&=\sqrt{\bar\alpha_{s-1}}\Big(\frac{x_s-\sqrt{1-\bar\alpha_s}\epsilon_{\theta}^{(s)}(x_s)}{\bar\alpha_t}\Big)\\
		&+\sqrt{1-\bar\alpha_{s-1}-\sigma^2_s}\cdot\epsilon^{(s)}_\theta+\sigma_s\epsilon_s,
		\label{eq:sample}
	\end{split}
\end{equation}
where $s=S,...,1$, and $\epsilon_\theta$ is estimated noise from $x_s$.

\section*{C. Technical Details}
\subsection*{C.1 Mathematical Symbols}
\label{appendix:math}

Key mathematical symbols used in the paper are documented in Table~\ref{tab:symbols}.

\begin{table}[htbp]
	\centering
	\caption{Key Mathematical Symbols}
	\label{tab:symbols}
	\small
	\setlength{\tabcolsep}{4pt}
	\renewcommand{\arraystretch}{1.2}
	\begin{tabular}{>{\centering\arraybackslash}m{0.28\columnwidth} >{\centering\arraybackslash}m{0.67\columnwidth}}
		\toprule
		\textbf{Symbol} & \textbf{Meaning}\\
		\midrule
		$w^t_\alpha$ & The smoke image at the $t$th frame and $\alpha$ viewing angle\\
		\midrule
		$w^t_{c,\alpha}$ & The clean image\\
		\midrule
		$w^t_{r,\alpha}$ & The rendered result for reconstructed density field\\
		\midrule
		$w^t_{f,\alpha}$ & The refined image\\
		\midrule
		$\alpha$ & $\alpha = \angle0^\circ$ for the input front view,\newline $\alpha = \angle90^\circ$ for the side view\\
		\midrule
		$I^t$ & The set of images from multiple views at the $t$th frame\\
		\midrule
		$\rho$ & Density field\\
		\midrule
		$\hat \rho$ & Advected density field\\
		\midrule
		$\rho_{r,c}$ & Coarse-grained reconstructed density field\\
		\midrule
		$\rho_{r,f}$ & Fine-grained reconstructed density field\\	
		\midrule
		$\mathbf{u}$ & Velocity field\\
		\midrule
		$\mathbf{u_r}$ & Reconstructed velocity field\\		
		\midrule
		$\rho_{in}$ & Inflow state\\
		\midrule
		$\mathcal{A}$ & Differentiable advection operator\\
		\midrule
		$\mathcal{R}$ & Differentiable rendering operator\\
		\midrule
		\text{SvDiff} & Side-view synthesizer based on diffusion models\\
		\midrule
		\text{NvRef} & Novel refinement module\\
		\midrule
		$\mathcal{G}^c_{\rho}$ & Coarse-grained density generator\\
		\midrule
		$\mathcal{G}^f_{\rho}$ & Fine-grained density generator\\
		\midrule
		$\mathcal{G}_u$ & Velocity generator\\
		\bottomrule
	\end{tabular}
\end{table}

\subsection*{C.2 Multi-frame Training Algorithm}

\begin{figure}[!ht] 
	\centering
    \includegraphics[width=1.0\linewidth]{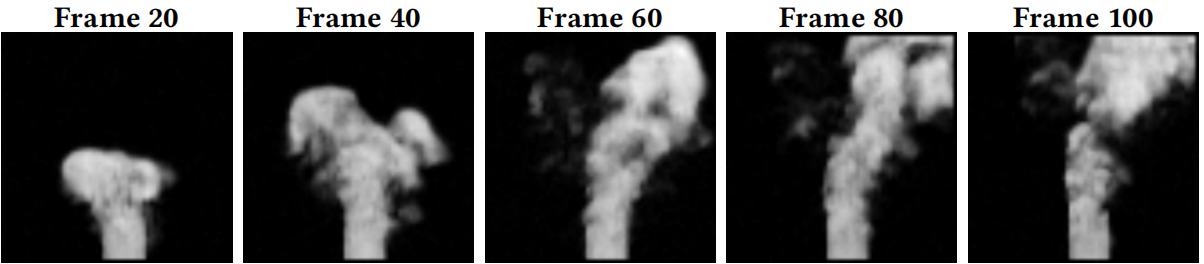}
	\caption{Side-view generation results affected by cumulative error.}
	\label{fig:dm_syn_woMul}
\end{figure}

If the previously synthesized frame is not used as one of the input conditions, the generated results exhibit significant cumulative errors, as shown in Fig.~\ref{fig:dm_syn_woMul}. To address this issue, we propose a multi-frame training algorithm, summarized in Alg.~\ref{alg:multiframe}, which incorporates the estimated clean image from the previous time step as a conditional input for the subsequent forward diffusion process.

\label{appendix:multi}
\begin{algorithm}[htbp]
	\caption{Multi-frame Training Algorithm for \text{SvDiff}.}
	\label{alg:multiframe}
	\begin{algorithmic}[1]
		\REQUIRE Number of iterations $it$, noise steps $S$, noise threshold $TQ$
		\REPEAT
		\STATE Sample $s \sim \text{Uniform}(\{1, \ldots, S\}$)
		\STATE $\rho^{t-1} = \mathcal{G}_\rho (w^{t-1}_{\angle 0^\circ}, w^{t-1}_{\angle 90^\circ})$
		\FOR{$i = 0, 1, 2, \ldots, it$}
		\STATE Condition $c^i: w^{i+t-2}_{c,\angle 90^\circ},\  w^{i+t-2}_{r,\angle 90^\circ},\ w^{i+t-1}_{c,\angle 90^\circ},\  w^{i+t-1}_{r,\angle 90^\circ},\ w^{i+t}_{\angle 0^\circ}$
		\STATE Clean image sample $x^i_0: w^{i+t}_{\angle 90^\circ}$
		\STATE Sample $\epsilon \sim \mathcal{N}(0, I)$
		\STATE $x^i_s = \sqrt{\bar{\alpha}_s} x^i_0 + \sqrt{1-\bar{\alpha}_s} \epsilon$
		\STATE $\hat \epsilon = \epsilon_\theta(x^i_s, c^i, s)$
		\STATE $\mathcal{L}_{noise} = \|\epsilon - \hat \epsilon\|^2$
		\IF{$s < TQ$}
		\STATE $\hat x^i_0 = \dfrac{x^i_s - \sqrt{1-\bar{\alpha}_s} \hat \epsilon}{\sqrt{\bar{\alpha}_s}}$
		\STATE $\rho^i_{r,c} = \mathcal{G}_\rho(w^{i+t}_{\angle 0^\circ}, w^{i+t}_{c,\angle 90^\circ})$
		\STATE $\mathbf{u}^{i-1} = \mathcal{G}_u(\rho^{i-1}, \rho^i_{r,c})$
		\STATE $w^{i+t}_{c,\angle 90^\circ} = \hat x^i_0, w^{i+t}_{r,\angle 90^\circ}=\mathcal{R}(\rho^i_{r,c}), \rho^{i-1} = \rho^i_{r,c}$
		\STATE $\mathcal{L}_{img} = \|x^i_0 - \hat x^i_0\|^2$
		\STATE $\mathcal{L}_{vel} = \|\nabla \cdot \mathbf{u}^{i-1}\|^2 +  \|\nabla \mathbf{u}^{i-1}\|^2$
		\STATE $\mathcal{L}_{sp} = \|H( w^{i+t}_{c,\angle90^\circ})-H(w^{i+t}_{\angle0^\circ})\|^2 $
		\ELSE 
		\STATE \textbf{break}
		\ENDIF
		\ENDFOR
		\STATE Take gradient step on $\mathcal{L}_{SvDiff}$
		\UNTIL{converged}
	\end{algorithmic}
\end{algorithm}

\subsection*{C.3 Progressive Refinement}

\begin{figure*}[ht]
	\centering
	\includegraphics[width=1.0\linewidth]{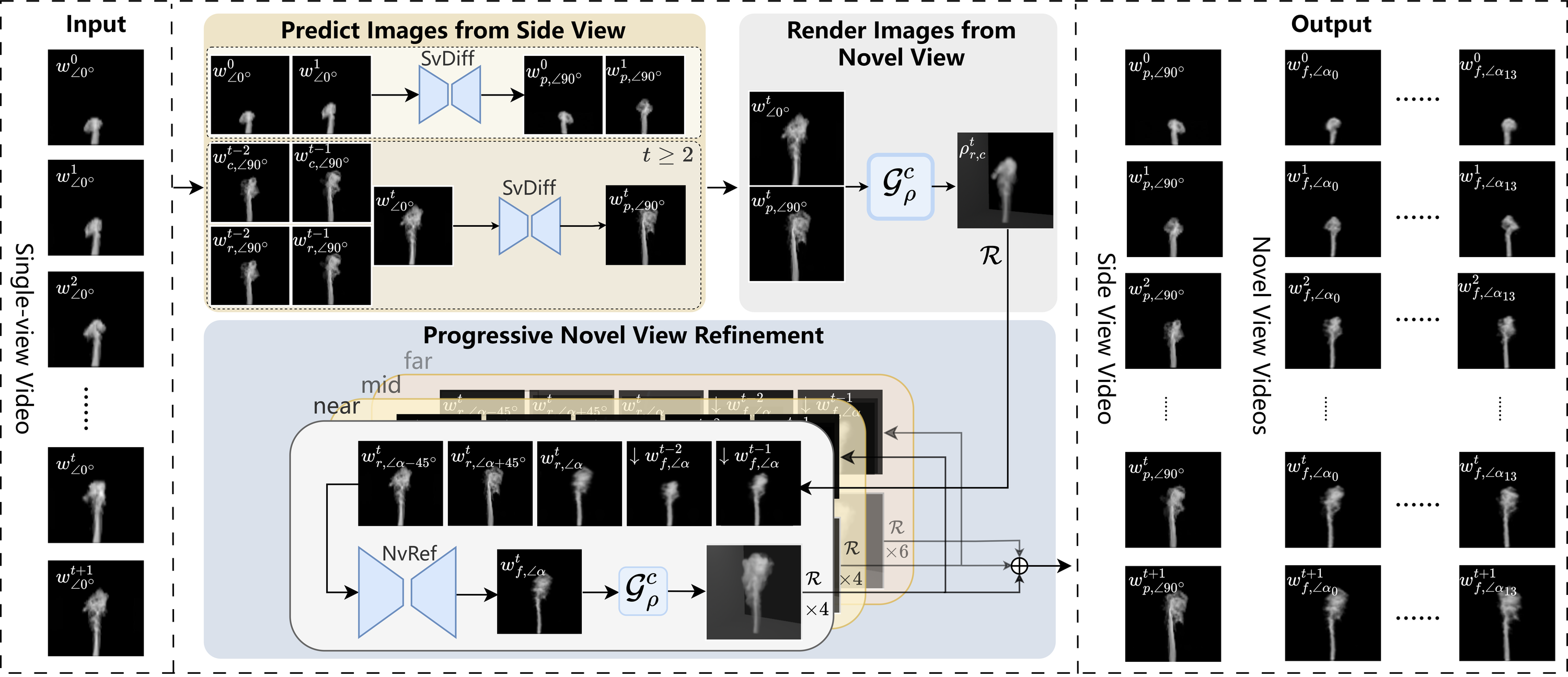}
	\caption{Procedure for side-view synthesis and novel view refinement. First, SvDiff predicts side-view images from input and previously generated images (when $t \geq 2$). Next, we reconstruct coarse density with $\mathcal{G}^c_\rho$ using front and side views, and render nearby novel views. Then, we iteratively refine novel views and reconstruct density, progressively extending from near to mid and far views, yielding multiple high-quality views for fine-grained reconstruction.}
	\label{fig:predict_and_SR}
\end{figure*}

\begin{figure}[ht] 
	\centering
    \includegraphics[width=1.0\linewidth]{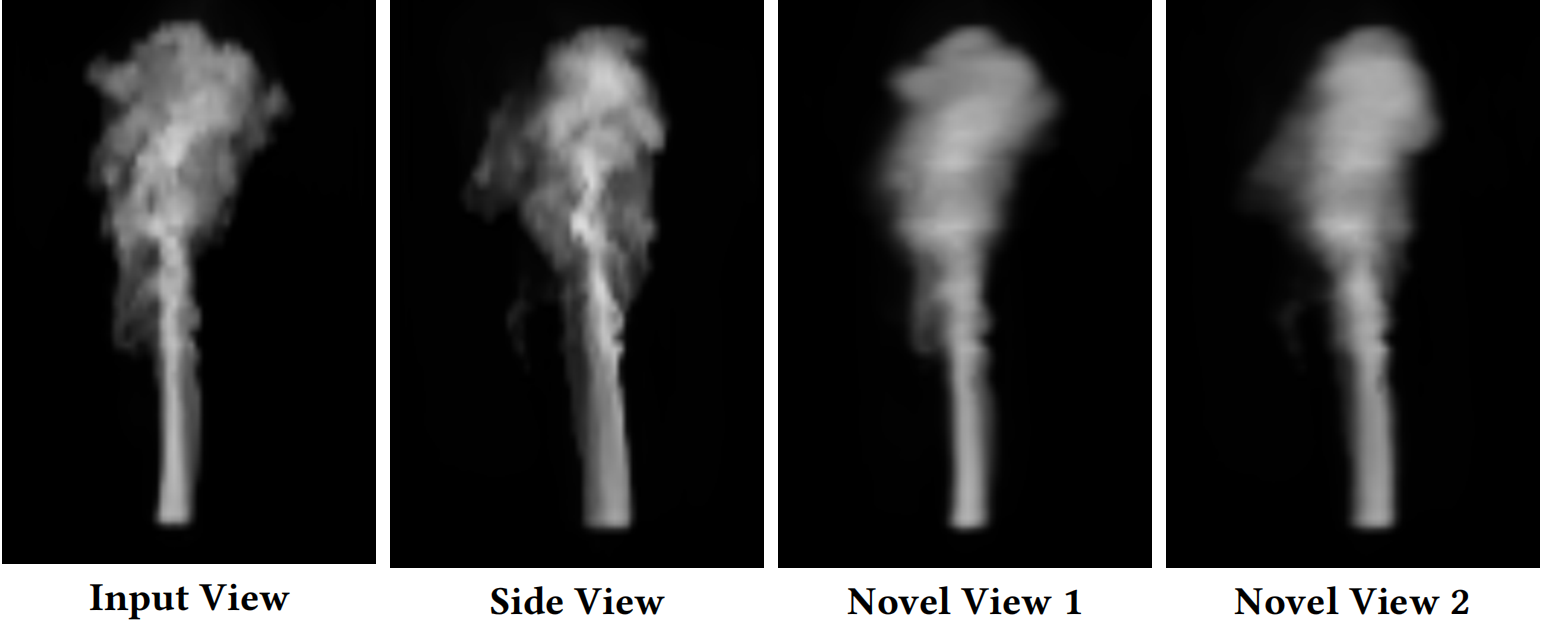}
	\caption{Rendering results of coarse-grained density field, which exhibits blurriness in novel views.}
	\label{fig:rho_c_sf}
\end{figure}

As shown in Fig.~\ref{fig:rho_c_sf}, $\rho^t_{r,c}$ appears blurry in novel views due to limited available information. To address this, we introduce a progressive refinement module that incrementally enhances the blurred novel images, improving clarity from near to far views, as summarized in Alg.~\ref{alg:nvref}.

\begin{algorithm}[t]
	\caption{Progressive Novel View Refinement.}
	\label{alg:nvref}
	\begin{algorithmic}[1]
		\REQUIRE Current frame $t$; coarse density $\rho_c^t$; near/mid/far view sets $nv$, $mv$, $fv$; angular offset $\beta$; refined images from previous frames $w_f^{t-1}$, $w_f^{t-2}$
		\STATE $\textit{ViewSets} \leftarrow \{nv,\ mv,\ fv\}$
		\FOR{each view set $V$ in $\textit{ViewSets}$}
		\STATE \textit{\# Rendering and refinement for the same view type}
		\FOR{each view angle $\alpha$ in $V$}
		\STATE $w^t_{r,\alpha} = \mathcal{R}(\rho_c^t,\,\alpha)$
		\STATE $w^t_{r,\alpha-\beta} = \mathcal{R}(\rho_c^t,\,\alpha-\beta)$
		\STATE $w^t_{r,\alpha+\beta} = \mathcal{R}(\rho_c^t,\,\alpha+\beta)$
		\STATE $w^t_{f,\alpha} = \mathrm{NvRef}\!\left(
		w^t_{r,\alpha-\beta} \oplus w^t_{r,\alpha+\beta} \oplus w^t_{r,\alpha}\right.$
		\STATE \hspace{2.5em} $\left.\oplus\, {\downarrow}\, w^{t-1}_{f,\alpha} \oplus {\downarrow}\, w^{t-2}_{f,\alpha} \right)$
		\ENDFOR
		\STATE \textit{\# Density reconstruction using all refined images obtained}
		\STATE $\rho_c^t = \mathcal{G}_\rho(\text{all refined imgs})$
		\ENDFOR
		\STATE \textit{\# After the final iteration}
		\STATE $\rho_f^t \leftarrow \rho_c^t$
	\end{algorithmic}
\end{algorithm}

\subsection*{C.4 Density Generator}
\label{appendix:density_generator}
\begin{figure}[ht]
	\centering
	\includegraphics[width=1\linewidth]{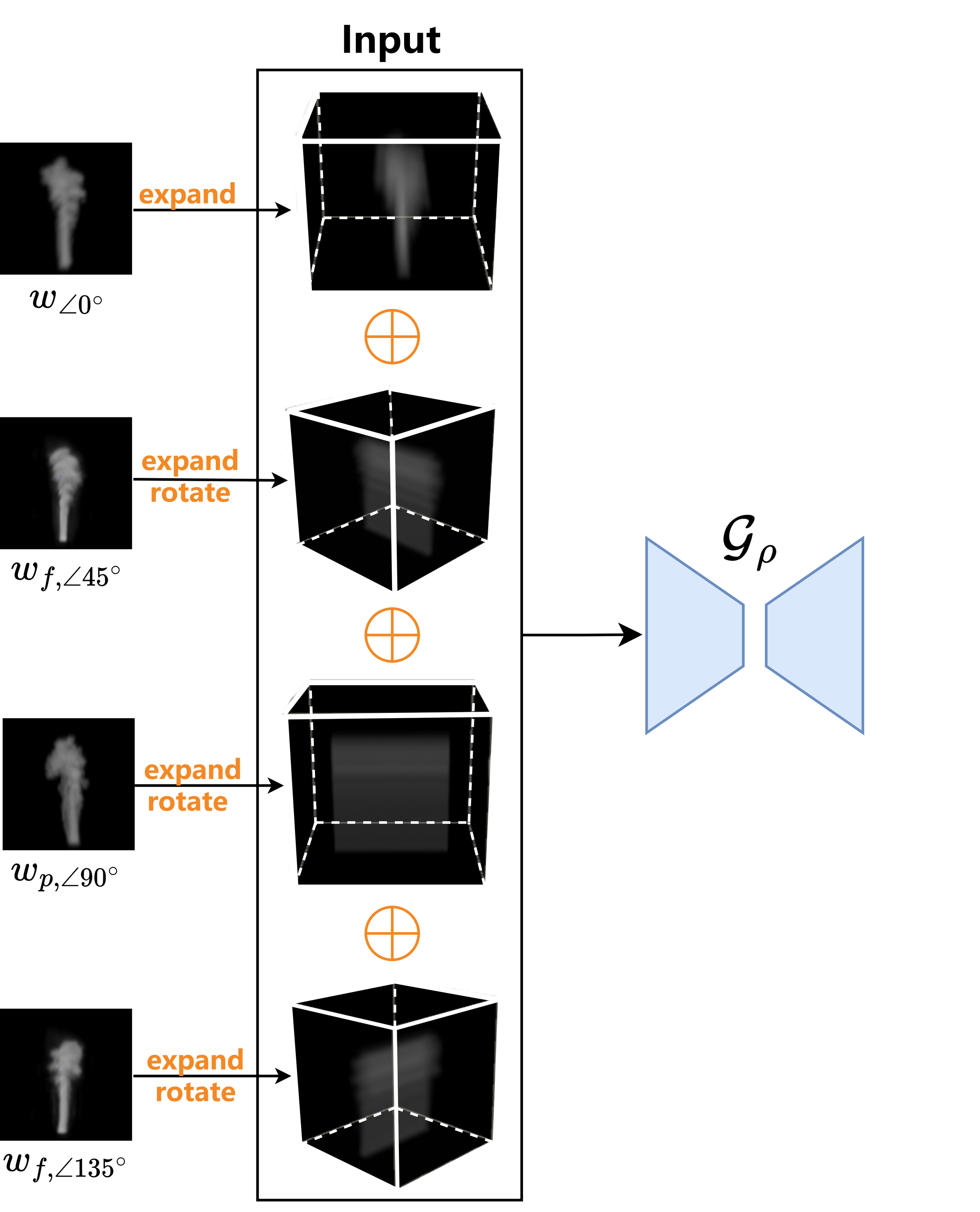}
	\caption{The architecture of density generator. The illustration depicts the case with four input images.}
	\label{fig:grho_model}
\end{figure}

To provide 3D input from 2D images, we transform the image through expansion to match the required dimensions, and concatenate them from multiple viewpoints, as shown in Fig.~\ref{fig:grho_model}. To be specific, $\mathcal{G}_{\rho}$ adopts the UNet3+ architecture with 3D convolutions. 

\subsection*{C.5 Velocity Estimation}
\label{subsec:velocity_estimation}

To reconstruct temporal and physically reasonable smoke dynamics, we establish a velocity generator $\mathcal{G}_u$ to estimate the velocity field based on two density fields of consecutive frames:
\begin{equation}
	\mathbf{u}_r^t=\mathcal G_u(\rho^t,\rho^{t+1}),
\end{equation}
which is supervised by $\mathcal{L}_u=\|\mathbf{u_r}-{\mathbf{u}}\|^2$. Additionally, to satisfy the divergence-free requirement in Eq.~\ref{eq:div}, we introduce another divergence loss as
$\mathcal{L}_{div}=\|\nabla \cdot \mathbf{u_r}-\nabla \cdot {\mathbf{u}}\|^2$.

To ensure long-term robustness and reduce the adverse impact of the reconstruction errors in density, we employ a differentiable advection operator $\mathcal{A}$ based on Eq.~\ref{eq:transport}, to formulate an advection loss term for the velocity generator. The advection operator $\mathcal{A}$ transports the density field $\rho$ based on the velocity field $\mathbf{u}$, expressed as:
\begin{equation}
	\hat \rho^{t} = \mathcal{A}(\rho^{t-1},\mathbf{u}_r^{t-1},\rho_{in},dt),
\end{equation}
where the density field obtained through velocity-based advection is called the advected density field, denoted as $\hat{\rho}$, $\rho_{in}$ is the dynamic inflow, and $dt$ is the time step. Similar to the density generator, we employ the following 3D density-based and 2D image-based advection loss terms:
\begin{align}
	\mathcal{L}_{advect}&=\lambda_{\hat{\rho}} \|\rho-\hat{\rho}\|^2 + \lambda_{\mathcal R} \| \mathcal R(\rho) -\mathcal R(\hat{\rho})\|^2.
\end{align}
Based on the advected density field $\hat{\rho}$, we modify the input of $\mathcal{G}_u$ to ensure that the velocity field can be corrected through the advected density field, with the formula being:
\begin{equation}
	\mathbf{u}_r^t=\mathcal G_u(\hat \rho^t,\rho^{t+1}).
\end{equation}

\subsection*{C.6 Inflow Estimation}
\label{subsec:inflow_estimation}
The inflow state has a tremendous impact on the visual pattern of smoke phenomena, which cannot be ignored in smoke reconstruction. In long-term evolution, underestimating the inflow will lead to an inability to fill the smoke volume in later time steps, while overestimating can cause obvious instability, ultimately failing to match the input images~\cite{eckert2019scalarflow}. 

To address this issue, we propose to estimate the inflow state frame-by-frame, that determines the inflow of current frame based on two adjacent density fields $\hat \rho^t$ and $\rho^{t+1}$, the velocity field $\mathbf{u}^t$, and the input image $w^{t+1}_{\angle0^\circ}$. Specifically, for each frame, we initialize a random smoke source $\rho_{in}$ and iteratively optimize the inflow source by minimizing the following loss function:
\begin{equation}
	\begin{split}
		\mathcal{L}_s&=
		\|\rho_r^{t+1}-\mathcal{A}(\hat{\rho}_r^t,\mathbf{u}_r^t,\rho_{in}^t,dt)\|^2\\&+\|w^{t+1}_{\angle0^\circ}-\mathcal{R}(\mathcal{A}(\hat{\rho}_r^t,\mathbf{u}_r^t,\rho_{in}^t,dt),\angle0^\circ)\|^2\\&+\|\rho_{in}^{t-1}-\rho_{in}^t\|^2.
		\label{eq:source}
	\end{split}
\end{equation}
Additionally, to prevent overestimation of the inflow source, we enforce zeroing out portions of the source that exceed a height threshold.

By incorporating the velocity and inflow estimation with density evolution~\cite{qiu2021rapid}, we can impose strong physical constraints to augment the temporal coherence and visual realism of SmokeSVD, thus effectively removing long-term flickers and non-physical artifacts in reconstructed smoke dynamics.

\section*{D. Implementation Details and Experimental Settings}
\label{subsec:implementation}

\paragraph{Implementation Details.} Our method is trained in two stages. In the first stage, we train \text{SvDiff} and \text{NvRef} based on the multi-frame training scheme to estimate clean images. We employ DDIM (Denoising Diffusion Implicit Models) sampling~\cite{song2020denoising} described in Eq.~\ref{eq:sample} to accelerate the sampling process. Simultaneously, we also train the density generator $\mathcal{G}_\rho$ and the velocity generator $\mathcal{G}_u$. Our density generator $\mathcal{G}_\rho$ outputs smoke density fields with resolutions of $64^3$ (for synthetic datasets) or $64 \times 112 \times 64$ (for real-world datasets). In the second stage, we fine-tune the velocity generator $\mathcal{G}_u$ based on the pre-trained density generator $\mathcal{G}_\rho$. All the aforementioned experiments were conducted on an NVIDIA GeForce RTX 3090 (24GB) GPU, while the performance was tested on an NVIDIA GeForce RTX 2080 Ti (11GB) GPU. Since optimization-based and neural radiance field (NeRF) methods require training for a few hours, far exceeding the minute-level time consumption of our proposed method, their specific time cost is not listed in the table.

\paragraph{Dataset.} Based on the Eulerian method~\cite{kim2008wavelet}, we generated the required synthetic dataset by randomly modifying the wind fields, thermal fields, and the size and position of inflow regions in the scenarios. A total of 100 scenarios were generated, with each scene containing 150 frames. Additionally, we used post-processed images from the first 20 scenes of the ScalarFlow dataset~\cite{eckert2019scalarflow} to train and evaluate our model.

\paragraph{Benchmarks.} We compared our method with existing techniques that accept single-view videos as input for 3D smoke reconstruction, selecting GlobTrans~\cite{franz2021global}, NGT~\cite{franz2023learning}, PICT~\cite{wang2024physics}, and PINF~\cite{chu2022physics} as benchmarks. In our experiments, we modified the inputs of PICT and PINF to support single-view video input. Among these methods, GlobTrans reconstructs 3D smoke based on direct optimization algorithms, while PICT and PINF are based on Neural Radiance Fields (NeRF). These methods all require optimization for individual scenario, resulting in expensive time consumption and re-optimization requirement when changing scenarios. In contrast, the NGT method uses a trained neural network to estimate a single motion of smoke, avoiding direct optimization of the entire scenario, thereby significantly improving reconstruction speed and applicability.

\paragraph{Evaluation Metric.} For image-related tasks (including novel view generation, refinement, and rendered images from reconstructed density fields), we use Mean Square Error (MSE), Root Mean Square Error (RMSE), Peak Signal-to-Noise Ratio (PSNR), Structural Similarity Index (SSIM)~\cite{wang2004image}, Fréchet Inception Distance (FID)~\cite{heusel2017gans}, Learned Perceptual Image Patch Similarity (LPIPS)~\cite{zhang2018unreasonable}, and STYLE similarity to measure the similarity between generated images and ground truth images. The STYLE similarity is defined as the $L1$ difference between the Gram matrices of features extracted from the generated results and the ground truth using VGG19. Additionally, we evaluate the feature consistency between generated images and ground truth images with $\mathcal{L}_{sp}$. For reconstruction tasks, we use RMSE of density fields, divergence and gradient of velocity fields to measure the similarity between reconstructed and ground truth physical fields.

\section*{E. More Evaluations}

\begin{figure*}[!ht] 
	\centering
    \includegraphics[width=1\linewidth]{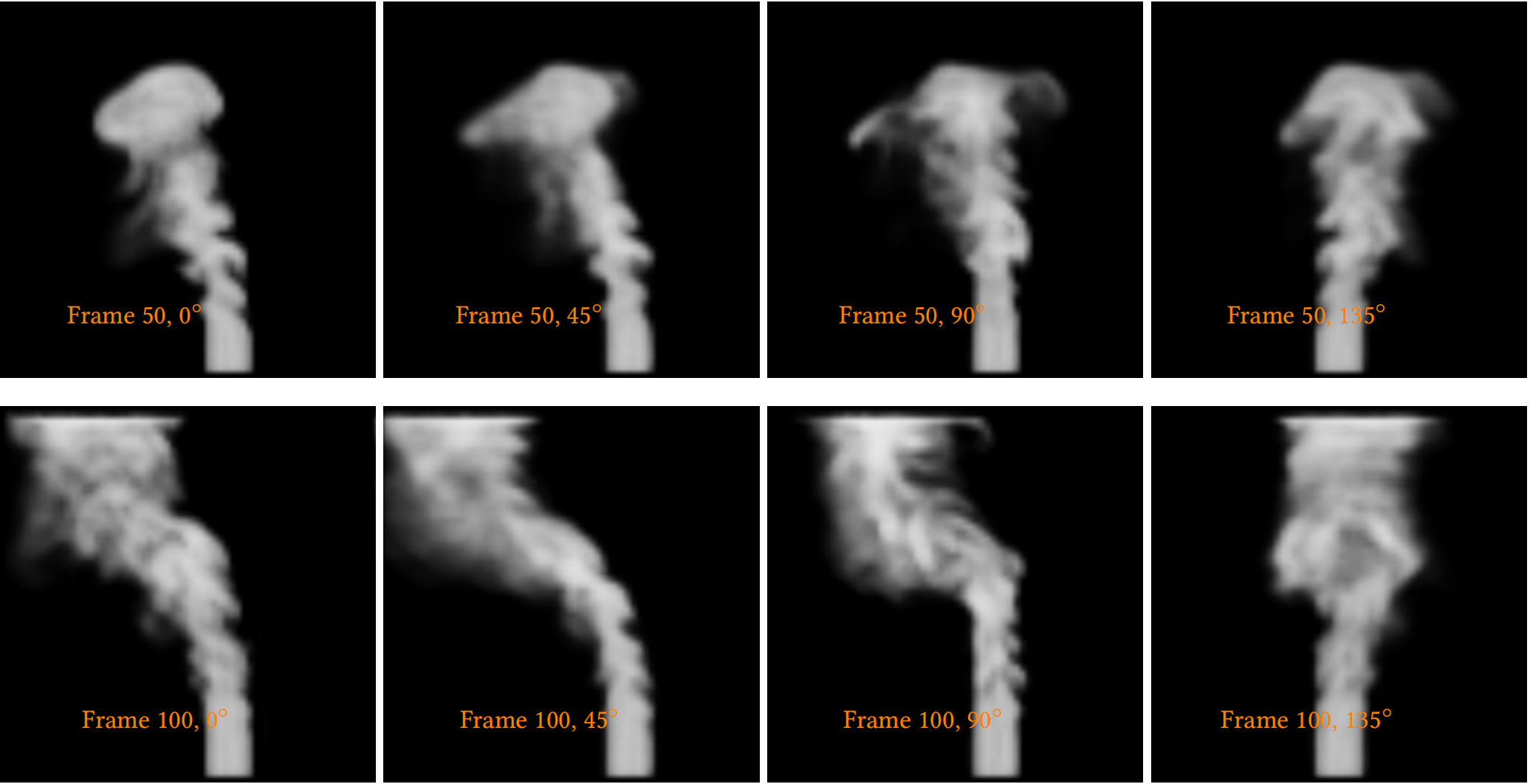}
	\caption{The rendering results of reconstructed density field at multiple views based on our proposed method.}
	\label{fig:our_syn_difAngle}
\end{figure*}

\paragraph{Results on Synthetic Dataset.} Fig.~\ref{fig:our_syn_difAngle} demonstrates the qualitative performance of our method on the synthetic dataset, where the density field resolution of the reconstructed scenario is $64^3$. By generating novel view images, our method significantly alleviates the ill-posed problem in single-view video based reconstruction, and the rendering results of reconstructed density fields perform well across different views. 

\paragraph{Side-View Quality.}  We employ optical flow analysis as temporal consistency metrics (Table~\ref{tab:optical_flow}). We achieve performance closest to GT (15min vs. GT's 30 hours): Max (2nd best) indicates minimal flickers, Avg shows reasonable dynamics comparable to NGT/GT, and Std validates consistency. Note that PICT's low metrics stem from depth-blur eliminating motion detail. 

\begin{table}[htb]
	\centering
	\vspace{-3mm}
	\caption{Optical flow statistics over 120 frames on ScalarFlow.}
	\vspace{-3mm}
	\label{tab:optical_flow}
	\resizebox{0.48\textwidth}{!}{
		\begin{tabular}{c|ccccccc}
			\toprule
			Metric & Reference & GT & NGT & PINF & PICT & FluidNexus & Ours \\
			\midrule
			Max. & 0.0896 & \textbf{0.0953} & 0.1272 & 0.1861 & 0.5890 & 0.2166 & \underline{0.1208} \\
			Avg. & 0.0593 & \underline{0.0639} & \textbf{0.0630} & 0.1274 & 0.0253 & 0.0765 & 0.0767 \\
			Std Dev  & 0.0091 & \underline{0.0121} & 0.0170 & 0.0185 & \textbf{0.0116} & 0.0348 & 0.0158 \\
			\bottomrule
		\end{tabular}
	}
	\vspace{-3mm}
\end{table}

\begin{figure*}[!ht] 
	\centering
	\includegraphics[width=1\linewidth]{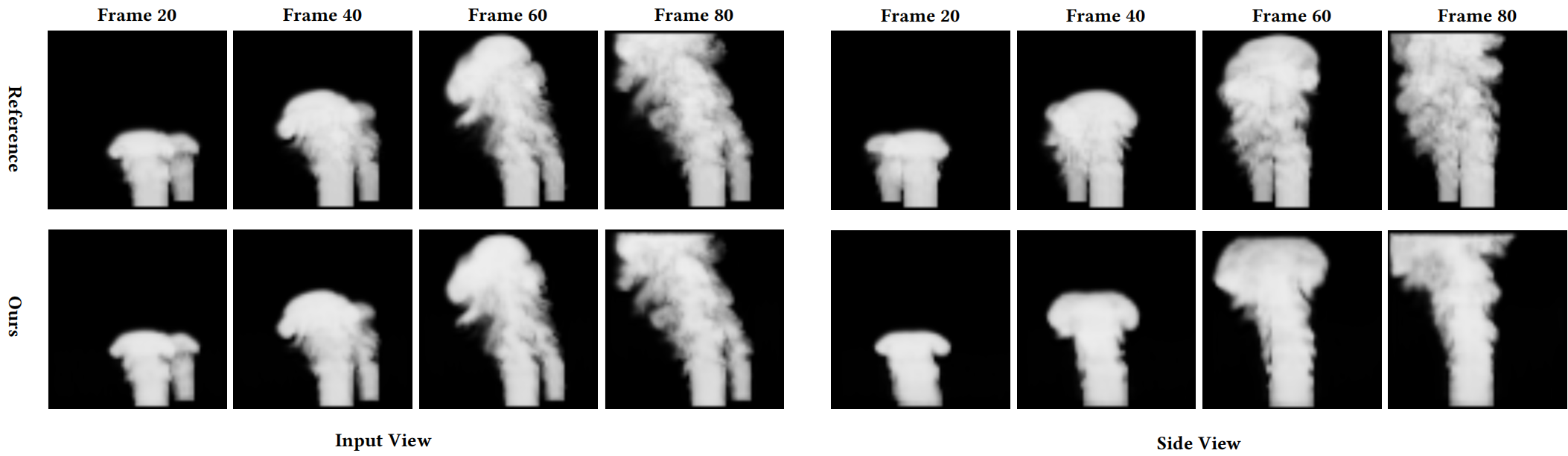}
	\caption{Reconstruction result for a multi-plume scenario, shown from both input and side views.}
	\label{fig:multi_plume}
\end{figure*}

\begin{figure*}[htb]
	\centering
	\includegraphics[width=1.0\linewidth]{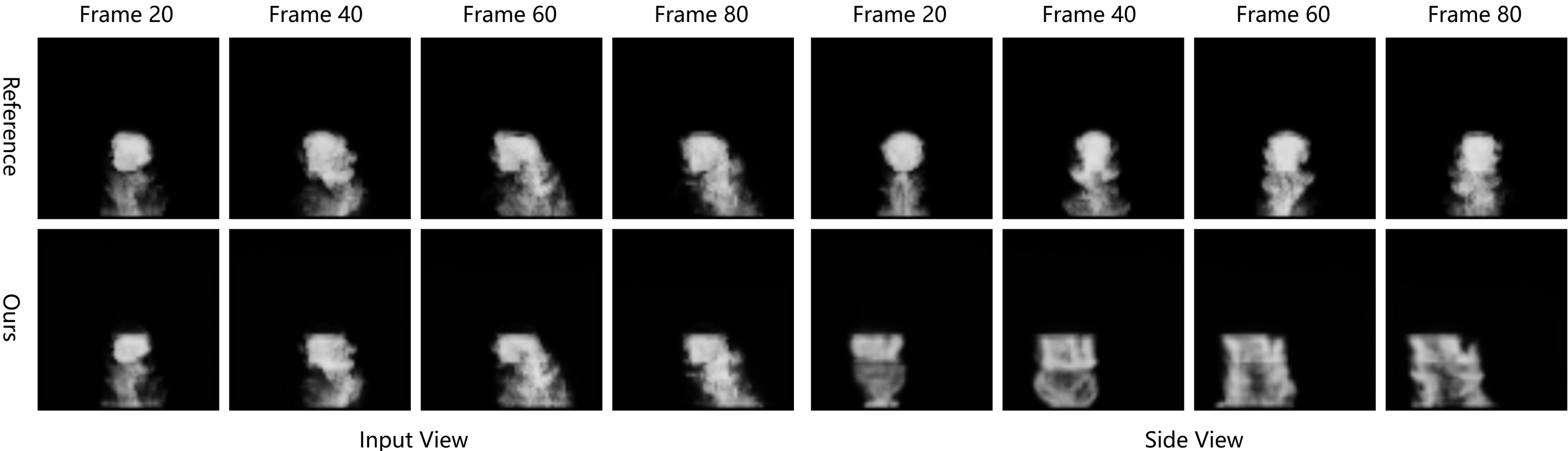}
	\caption{Reconstruction result for a dry ice scenario.}
	\label{fig:co2}
\end{figure*}

\paragraph{More Generalization Performance.}
We also test with multi-plume collisions and dry ice, as shown in Figs.~\ref{fig:multi_plume} and~\ref{fig:co2}. Our method performs well on various smoke shapes, which are fundamentally different from the single-source smoke scenes in our training dataset.

\begin{figure*}[!ht] 
	\centering
    \includegraphics[width=1\linewidth]{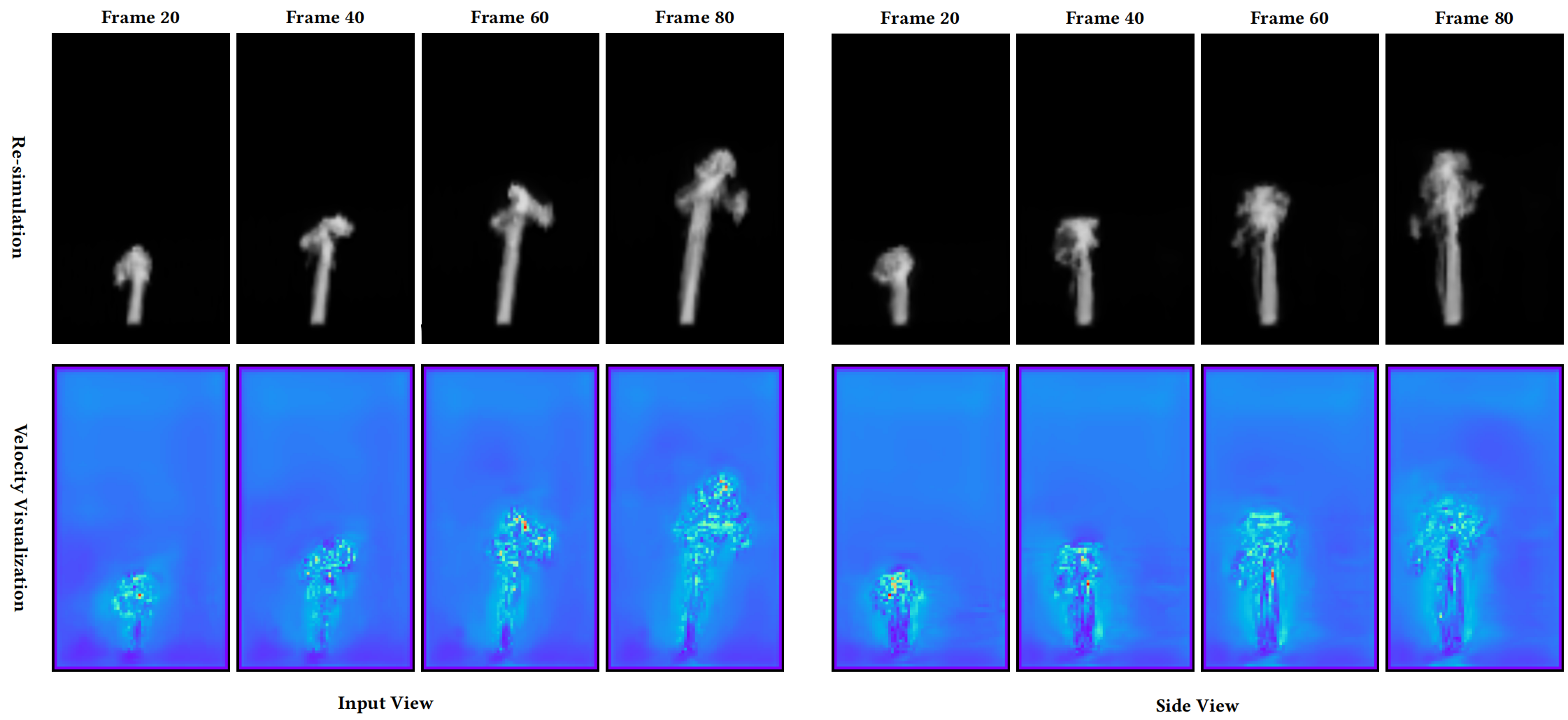}
	\caption{The rendered re-simulation results and velocity estimation visualization at the input view and the side view.}
	\label{fig:re-simulation}
\end{figure*}

\begin{figure*}[!ht] 
	\centering
    \includegraphics[width=1\linewidth]{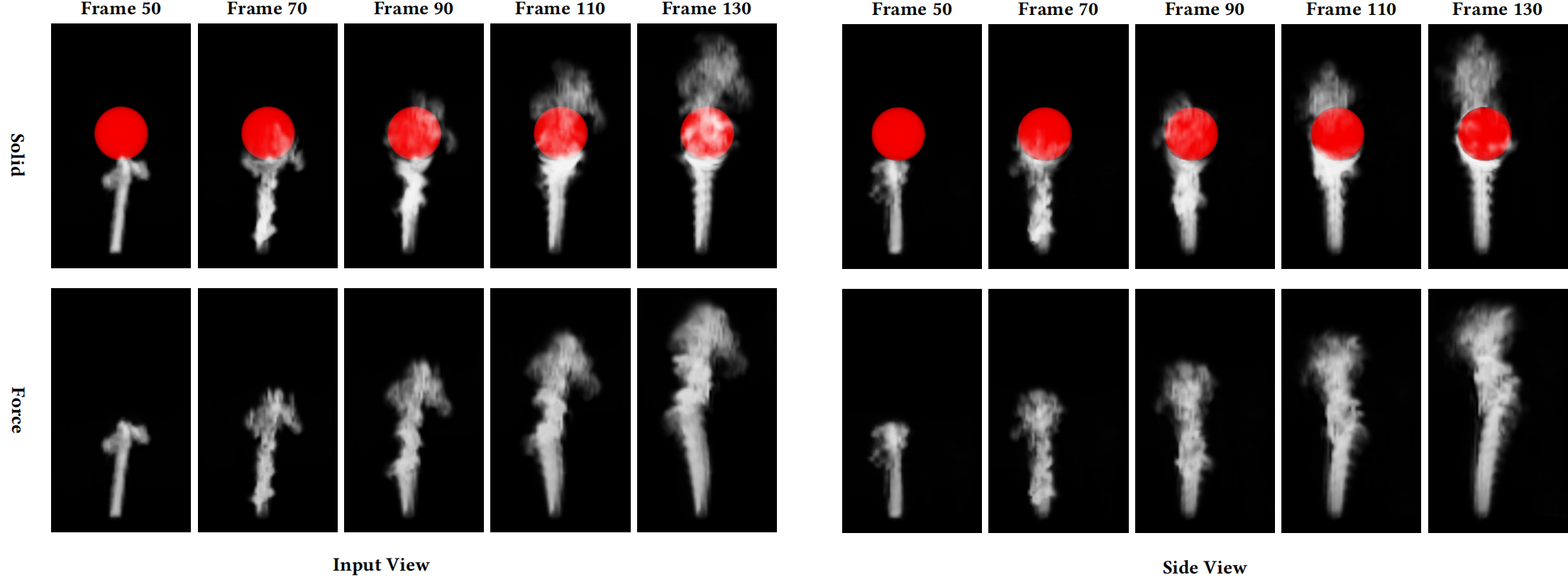}
	\caption{The re-simulation result with added fluid-solid coupling (top row), where we place a sphere obstacle (the red circle) at the $50$th time step, and external force field (bottom row).}
	\label{fig:interactive_simulation}
\end{figure*}

\paragraph{Interactive Simulation.}
Our reconstructed physical fields enable the re-simulation of input videos, and the generation of new smoke phenomena with controllable effects and enhanced detail, as shown in Figs.~\ref{fig:re-simulation} and~\ref{fig:interactive_simulation}. In Fig.~\ref{fig:interactive_simulation}, we demonstrate re-simulation results in which a newly added spherical obstacle (top row) or external force field (bottom) is introduced by projecting the reconstructed velocity field onto a new simulation domain.

\begin{figure*}[htb]
	\centering
	\includegraphics[width=1\textwidth]{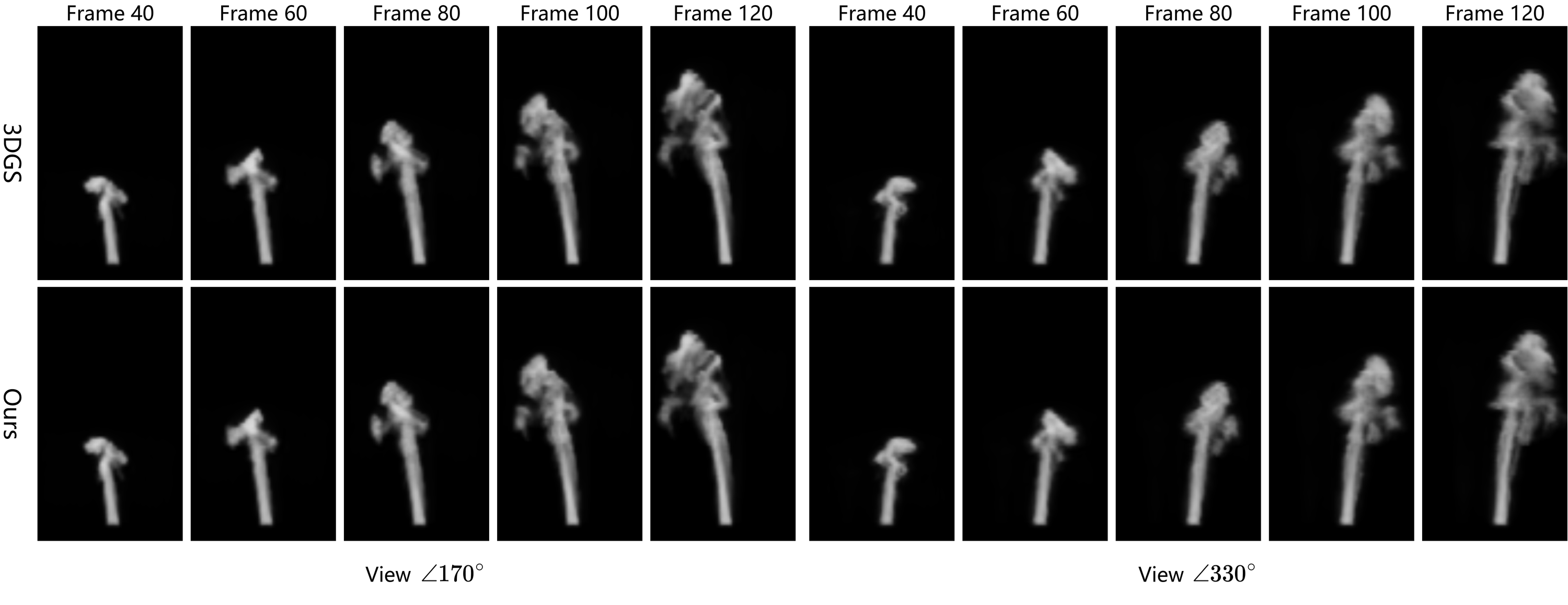}
	\caption{3DGS results (top) based on our synthesized novel views (bottom).}
	\label{fig:3dgs}
\end{figure*}

\begin{figure*}[htb]
	\centering
	\includegraphics[width=1\textwidth]{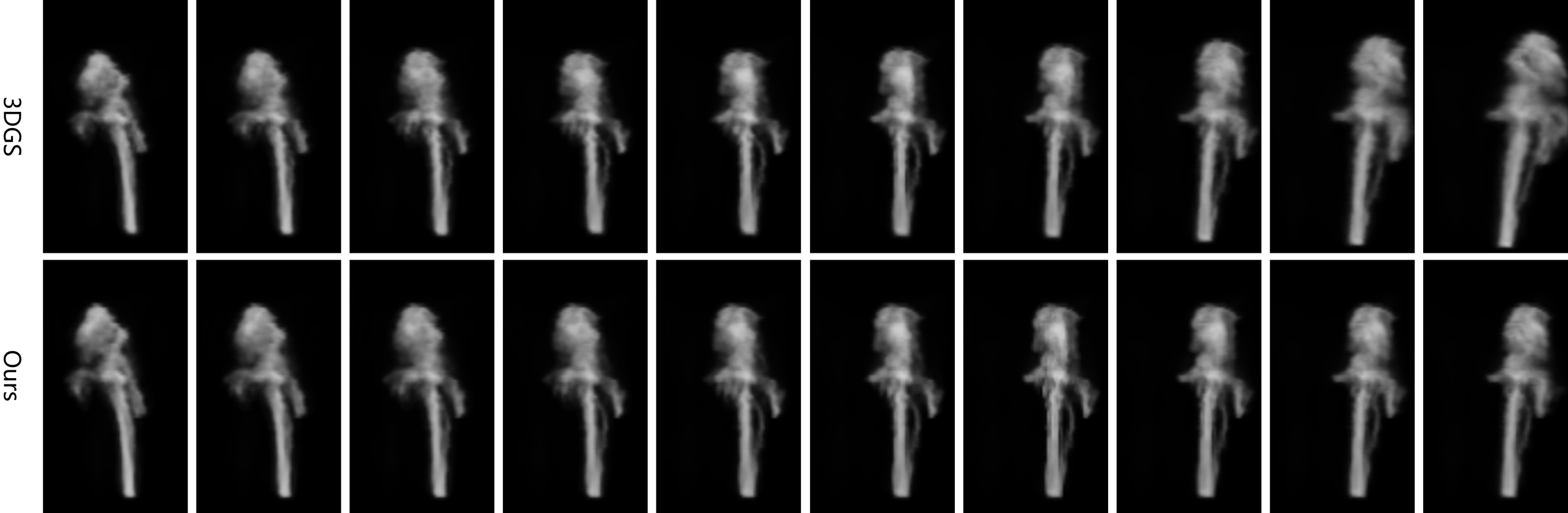}
	\caption{3DGS results (top) and our reconstruction result (bottom) under rotating views from \(\angle 210^\circ\) to \(\angle 300^\circ\).}
	\label{fig:3dgs2}
\end{figure*}

\paragraph{Compatibility with 3D Gaussian Splatting.}
Once sufficient novel views have been generated, our method can be seamlessly integrated with downstream applications such as 3D Gaussian Splatting (3DGS). As shown in Figs.~\ref{fig:3dgs} and~\ref{fig:3dgs2}, thanks to the multi-view consistency and well-structured spatiotemporal features provided by our approach, 3DGS is able to reproduce physically and visually plausible smoke sequences without the need for additional temporal processing.

\section*{F. More Ablation Studies}
\label{appendix:ablation}

\begin{figure}[htb] 
	\centering
    \includegraphics[width=1\linewidth]{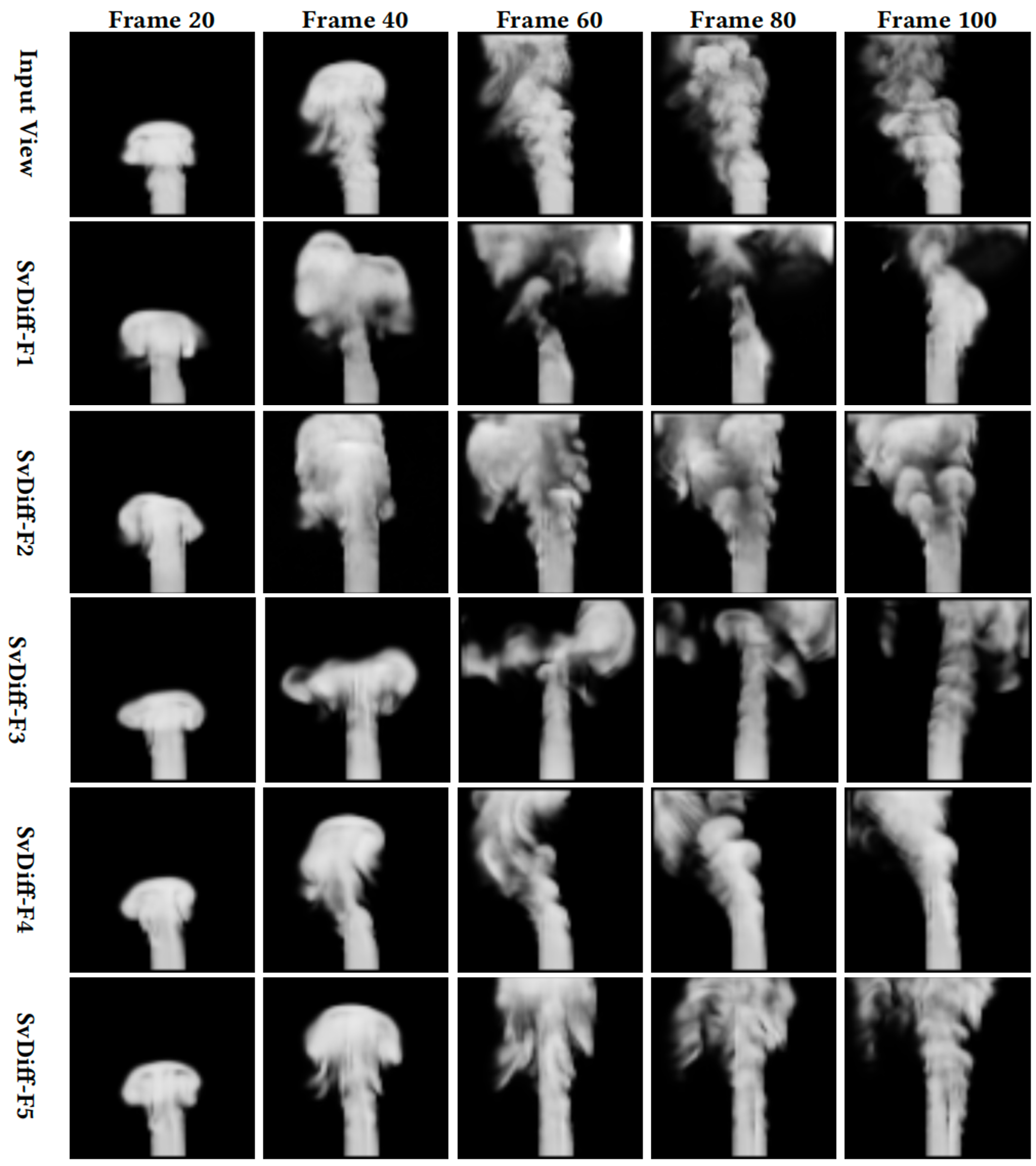}
	\caption{Qualitative comparison of side view synthesis with different frame numbers on the synthetic dataset.}
	\label{fig:ddim_syn}
\end{figure}

\paragraph{Effect of Frame Numbers.} We adopted a multi-frame training strategy to train the side-view synthesizer (\text{SvDiff}) and the novel view refinement module (\text{NvRef}). Taking \text{SvDiff} as an example, in the early stages of training, We fed \text{SvDiff} one image for a single forward diffusion process; subsequently, we gradually increased the number of training frames and forward diffusion times until the synthesis quality met the expectation. To determine the final number of training frames and forward diffusion timesteps, we tested different hyperparameter settings for \text{SvDiff}. Since the number of training frames equals the number of forward diffusion times, we named these hyperparameter settings based on the number of frames (e.g., SvDiff-F1, SvDiff-F2), as shown in Fig.~\ref{fig:ddim_syn}. As the number of training frames increased, the synthetic results gradually became more reasonable. For example, the SvDiff-F1 in Fig.~\ref{fig:ddim_syn} did not use the multi-frame information to estimate clean images, so due to the cumulative error, subsequent synthetic frames gradually deviated from reasonable smoke appearance. According to the results in Table~\ref{tb:nvgm_ddim}, we found that the \text{SvDiff} based on four forward diffusions (SvDiff-F4) achieves the best. Both qualitative and quantitative evaluations indicate that the multi-frame training strategy based on estimated clean images plays a crucial role in the long-term generation process of diffusion models.

\begin{table}[htb]
	\centering
	\caption{Quantitative comparison of \text{SvDiff} with different frame numbers on the synthetic dataset. We report $\mathcal{L}_{sp}$, LPIPS, and SSIM to measure the differences between synthetic images and reference images, and warp error to measure pixel-level distortion between consecutive frames based on mean squared error (MSE).}
	\resizebox{0.5\textwidth}{!}{
		\begin{tabular}{c|c c c c} 
			\hline 
			Algorithm & \(\mathcal{L}_{sp}\)$\downarrow$ & Warp Error$\downarrow$ & LPIPS$\downarrow$ & SSIM$\uparrow$ \\
			\hline
			reference & / & 0.0981 & / & / \\
			\hline
			SvDiff-F1 & 1.2601 & 0.2003 & 0.3873 & 0.4364 \\
			SvDiff-F2 & 1.2673 & 0.1819 & 0.3742 & \underline{0.5077} \\
			SvDiff-F3 & 1.0422 & \pmb{0.0915} & 0.3910 & 0.4997 \\
			SvDiff-F4 & \pmb{0.3475} & 0.1481 & \pmb{0.3384} & \pmb{0.5729} \\
			SvDiff-F5 & \underline{0.7081} & \underline{0.1259} & \underline{0.3779} & 0.5052 \\
			\hline
		\end{tabular}
	}
	\label{tb:nvgm_ddim}
\end{table}

\paragraph{Effect of View Numbers.} Our density generator can accept up to 16 smoke images from different viewpoints, with these views evenly distributed along a $180^\circ$ arc. To determine the optimal number of input views for fine-grained density reconstruction, we trained several density generators using 2, 4, 8, and 16 input images (denoted as $2-, 4-, 8-, 16-\mathcal{G}\rho$), and evaluated their performance. The quantitative results are presented in Table~\ref{tb:grho}.
In the experiment, when the number of input images was less than 16, images from other novel views were masked. All image metrics were evaluated based on 16 real viewpoints, and the quantitative analysis indicates that as the number of input views increases, the reconstruction quality gradually improves.
Therefore, in the coarse-grained density reconstruction stage, we used only a subset of views as input, whereas in the fine-grained stage, all 16 input views were utilized to provide richer information for high-quality reconstruction.

\begin{figure*}[htb] 
	\centering
    \includegraphics[width=0.8\linewidth]{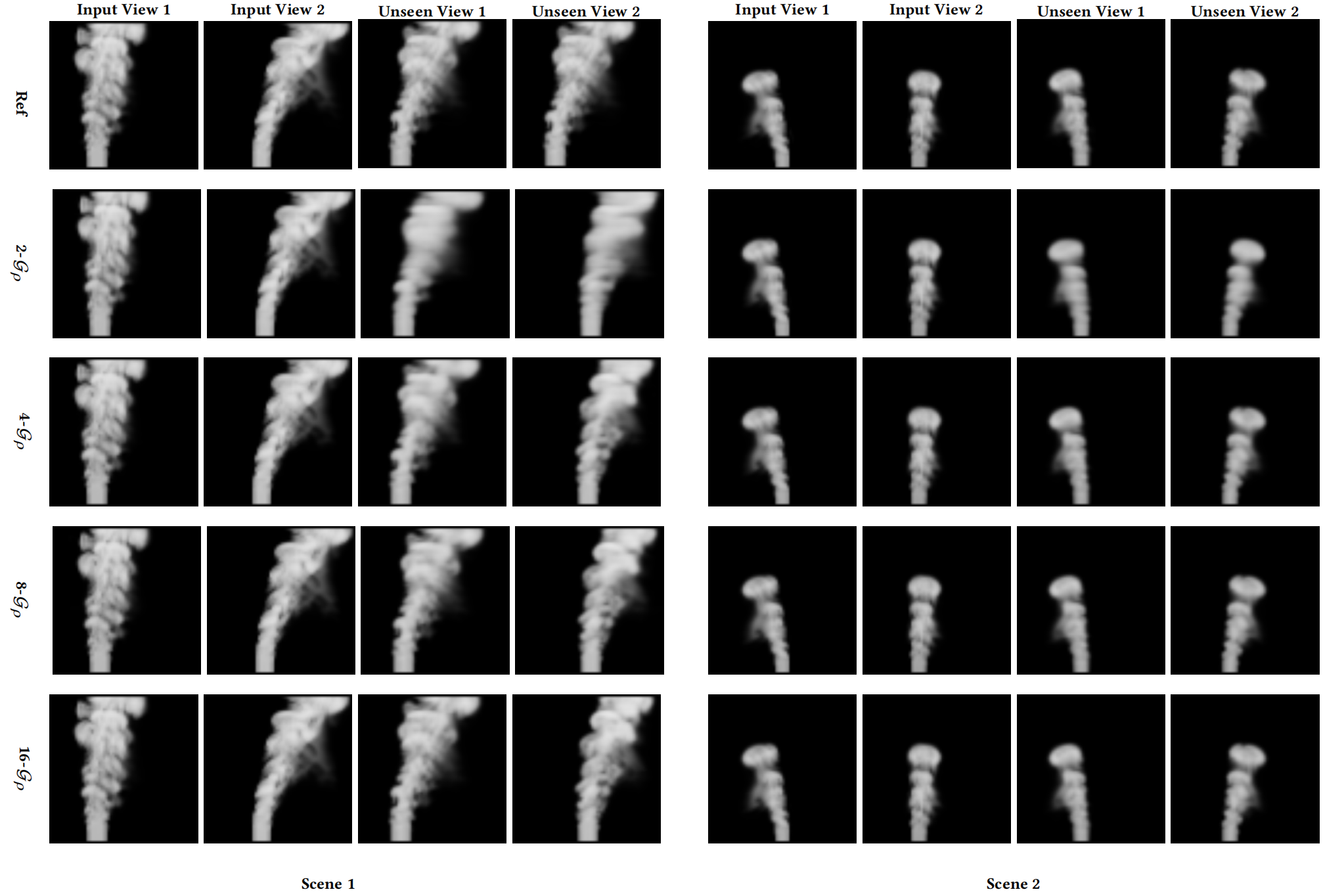}
	\caption{Qualitative comparison of density generators with different numbers of views on the synthetic dataset.}
	\label{fig:grho_syn}
\end{figure*}

\begin{table}[htb]
	\centering
	\caption{Quantitative evaluation of density generators with different numbers of input views on the synthetic dataset. The last five metrics are evaluated based on images from 16 views.}
	\resizebox{0.5\textwidth}{!}{
		\begin{tabular}{c|c|c c c c c} 
			\hline  
			View Num & \(\rho\) RMSE$\downarrow$ & RMSE $\downarrow$& SSIM$\uparrow$ & PSNR$\uparrow$ & LPIPS$\downarrow$ & FID$\downarrow$\\  
			\hline
			2 & 0.0356 & 0.0206 & 0.9795 & 37.0561 & 0.0417 & 31.0919 \\
			4 & 0.0256 & 0.0100 & 0.9915 & 43.1682 & 0.0205 & 9.7665\\
			8 & \underline{0.0186} & \underline{0.0058} & \underline{0.9960} & \underline{47.2533} & \underline{0.0099} & \underline{2.5882}\\
			16 & \pmb{0.0148} & \pmb{0.0043} & \pmb{0.9974} &\pmb{49.6970} &\pmb{0.0050}  &  \pmb{1.3745}\\
			\hline
		\end{tabular}
	}
	\label{tb:grho}
\end{table}

\begin{figure}[htb]
	\centering
	\includegraphics[width=1\linewidth]{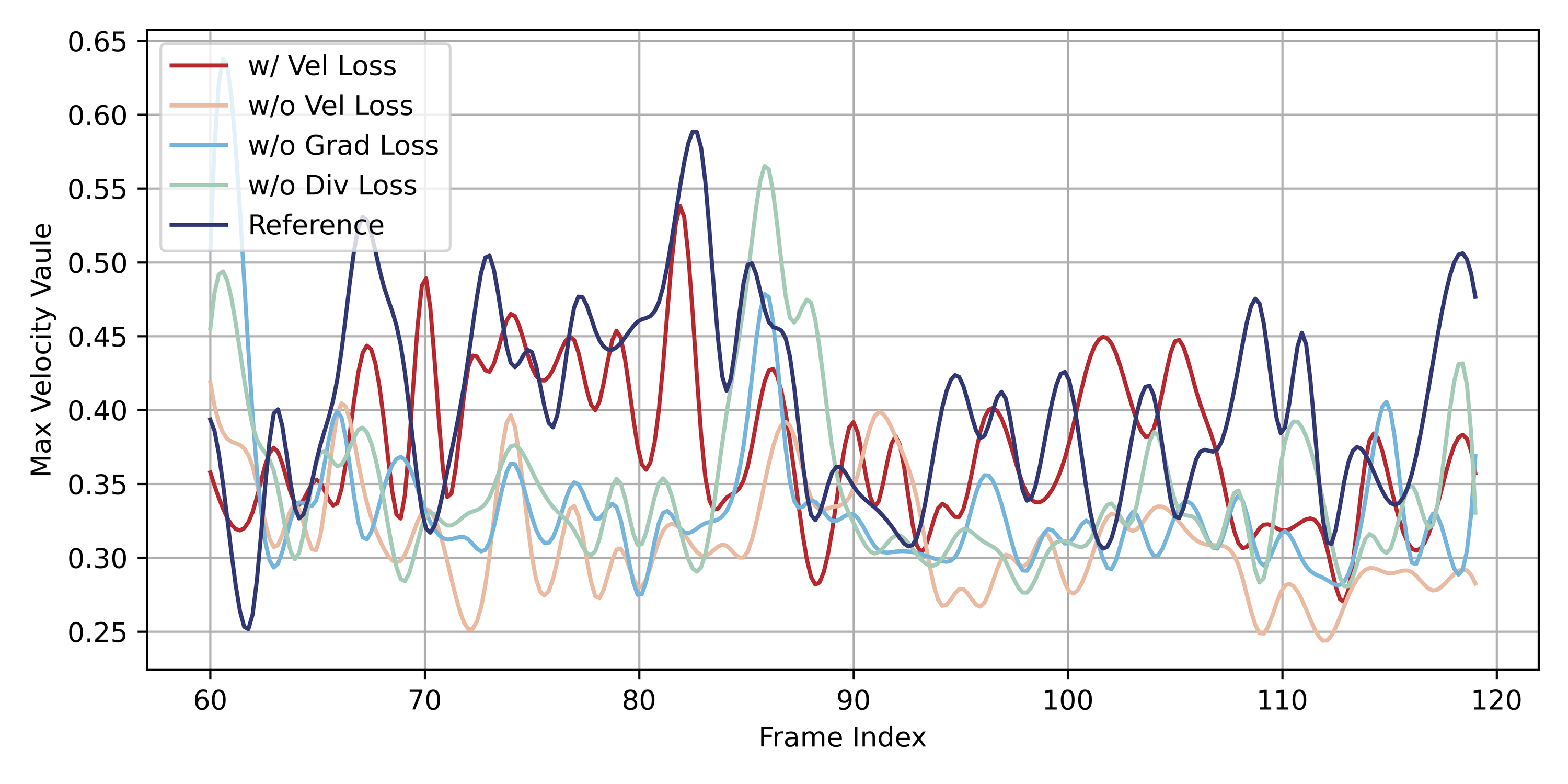}
	\caption{Comparison of the maximum values of reconstructed velocity fields by SvDiff with different loss functions at various time steps.}
	\label{fig:ablation_svdiff_newvelmax}
\end{figure}

\begin{figure}[htb]
	\centering
	\includegraphics[width=1\linewidth]{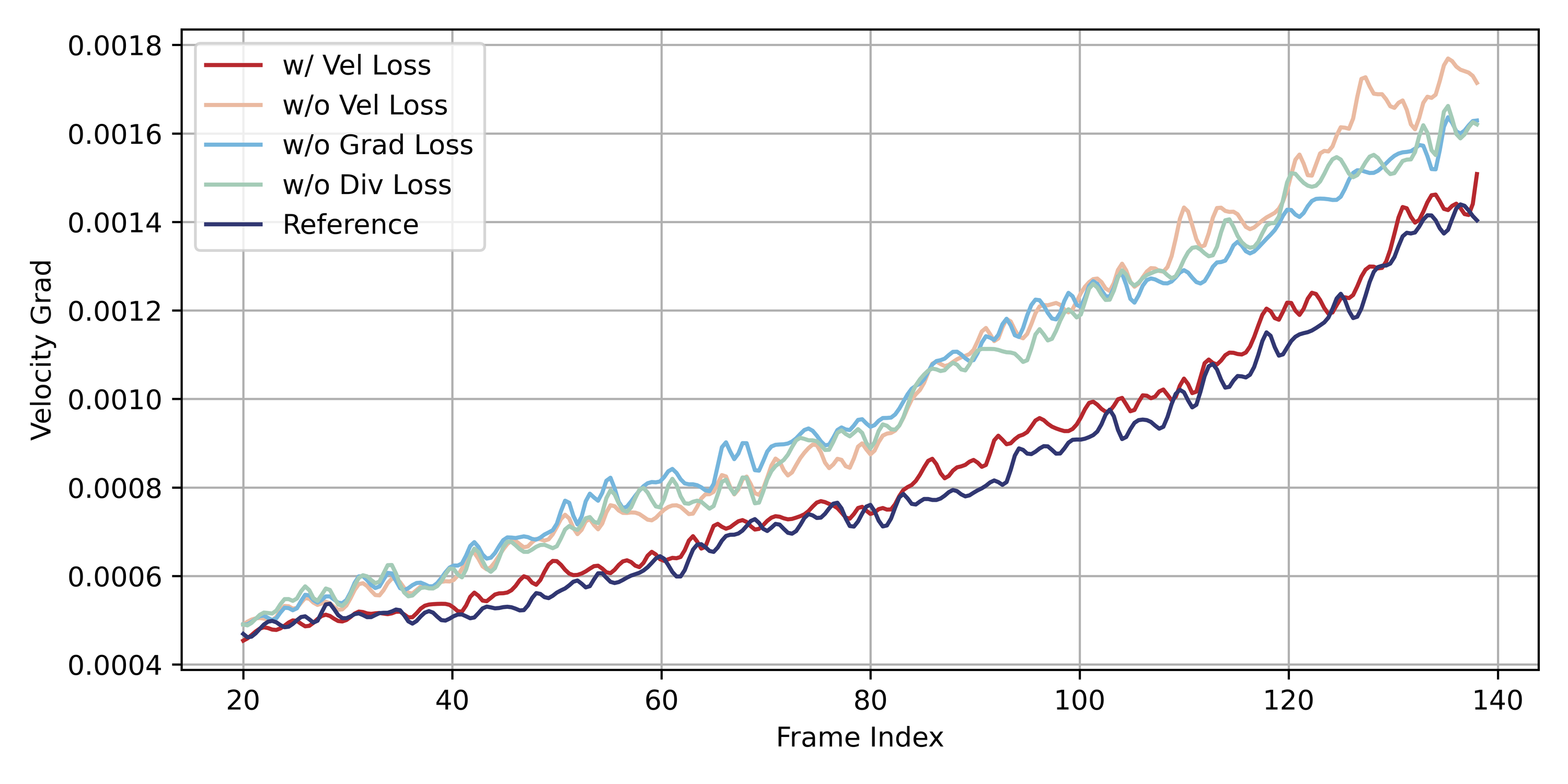}
	\caption{Comparison of the gradient of reconstructed velocity fields by SvDiff with different loss functions at various time steps.}
	\label{fig:ablation_svdiff_velgrad}
\end{figure}

\paragraph{Ablation on Side-view Synthesizer.}
We also visualized the maximum values and gradient of reconstructed velocity fields in Figs.~\ref{fig:ablation_svdiff_newvelmax} and~\ref{fig:ablation_svdiff_velgrad}. 

\begin{figure}[htb]
	\centering
	\includegraphics[width=1\linewidth]{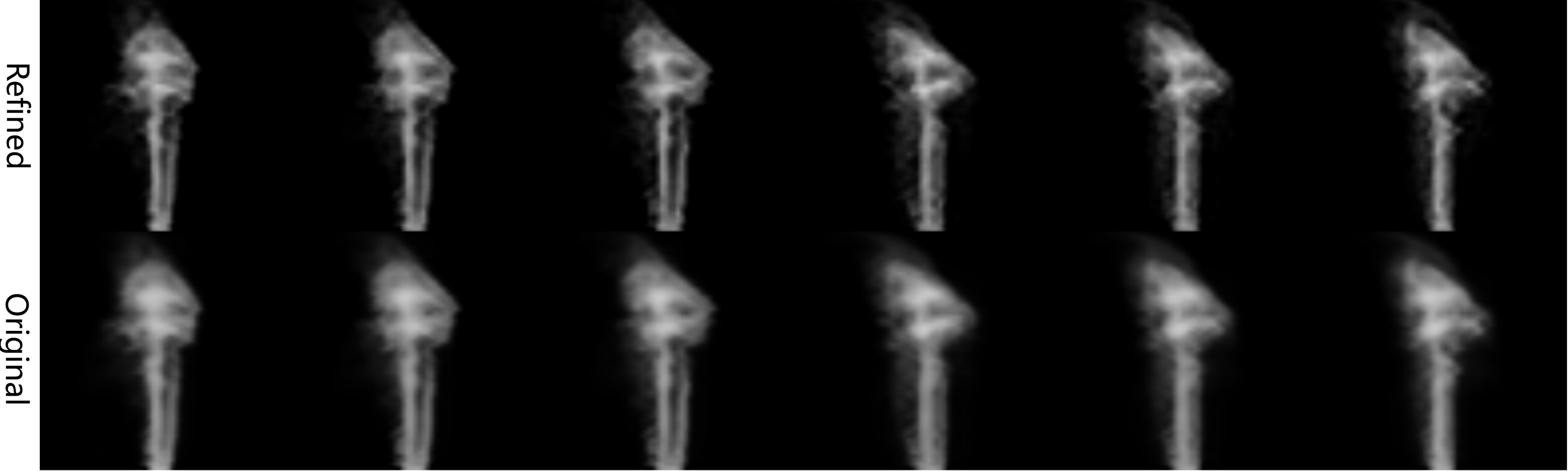}
	\caption{NGT combined with our refinement model.}
	\label{fig:NGTplusrefine}
\end{figure}

\begin{figure}[htb]
	\centering
	\includegraphics[width=1\linewidth]{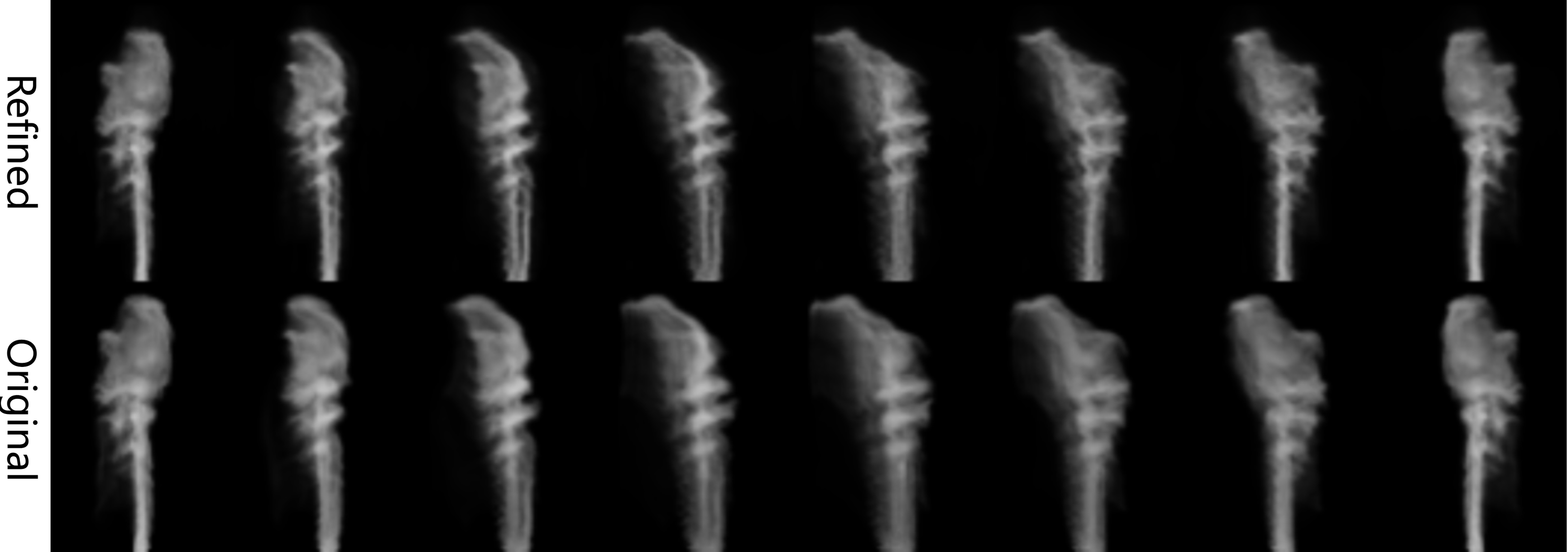}
	\caption{NGT combined with our reconstruction model.}
	\label{fig:NGTplusreconstruction}
\end{figure}

\paragraph{Ablation on Key Components.} Figs.~\ref{fig:NGTplusrefine} and~\ref{fig:NGTplusreconstruction} show NGT combined with our refinement and reconstruction. Our approach is compatible with NGT and further enhances its results, achieving high-quality reconstruction.

\section*{G. Limitation and Discussion}
\label{subsec:limitation}
While our proposed framework demonstrates strong performance in reconstructing dynamic smoke from single-view input, several limitations remain. First, the current method assumes a relatively clean background and consistent lighting conditions; in real-world scenarios with complex backgrounds or varying illumination, the quality of side-view synthesis and subsequent reconstruction may degrade. Second, although our progressive refinement strategy improves multi-view consistency, the approach still relies on the accuracy of the initial side-view synthesis, significant errors in early stages can propagate and affect the final results. Third, our model is primarily evaluated on synthetic and controlled real-world datasets; its generalization to highly diverse or outdoor smoke phenomena remains to be further validated. Additionally, the computational cost, while lower than optimization-based methods, can still be significant when scaling to higher resolutions or longer sequences. Finally, our framework currently focuses on grayscale smoke and does not explicitly handle colored smoke, solid obstacles, or interactions with complex environments. Future work could address these limitations by incorporating more robust background modeling, exploring domain adaptation techniques, extending the framework to handle color and multi-phase flows, and integrating more advanced physical constraints to further enhance realism and generalization.

{
    \small
    \bibliographystyle{ieeenat_fullname}
    \bibliography{main}
}